\documentclass[reprint,aps,prx,letterpaper,superscriptaddress]{revtex4-2}
\usepackage{amssymb}
\usepackage{amsmath}
\usepackage{graphicx}
\usepackage{footnote}
\usepackage{booktabs}
\newcommand{\ra}[1]{\renewcommand{\arraystretch}{#1}}
\usepackage[usenames,dvipsnames,svgnames,table]{xcolor}
\usepackage[T1]{fontenc}
\usepackage[
pdfauthor={Paul Nation, Abdullah Ash Saki, Sebastian Brandhofer, Luciano Bello, Shelly Garion, Matthew Treinish, Ali Javadi-Abhari},
pdftitle={Benchmarking the performance of quantum computing software},
bookmarks=true,colorlinks=true,linkcolor=OrangeRed,urlcolor=RoyalBlue,citecolor=DarkOrchid]{hyperref}

\usepackage{xr}
\begin{document}

\title{Benchmarking the performance of quantum computing software}

\author{Paul D. Nation}
\email[E-mail: ]{paul.nation@ibm.com}
\author{Abdullah Ash Saki}
\affiliation{IBM Quantum, IBM T. J. Watson Research Center, Yorktown Heights, NY, 10598 USA}
\author{Sebastian Brandhofer}
\affiliation{IBM Quantum, IBM Germany Research \& Development, B\"{o}blingen Germany}
\author{Luciano Bello}
\affiliation{IBM Quantum, IBM Research Europe, Zurich Switzerland}
\author{Shelly Garion}
\affiliation{IBM Quantum, IBM Research Israel, Haifa 3498825, Israel}
\author{Matthew Treinish}
\author{Ali Javadi-Abhari}
\affiliation{IBM Quantum, IBM T. J. Watson Research Center, Yorktown Heights, NY, 10598 USA}

\begin{abstract}
We present Benchpress, a benchmarking suite for evaluating the performance and range of functionality of multiple quantum computing software development kits. This suite consists of a collection of over $1000$ tests measuring key performance metrics for a wide variety of operations on quantum circuits comprised of up to $930$ qubits and $\mathcal{O}(10^{6})$ two-qubit gates, as well as an execution framework for running the tests over multiple quantum software packages in a unified manner. We give a detailed overview of the benchmark suite, its methodology, and generate representative results over seven different quantum software packages. The flexibility of the Benchpress framework allows for benchmarking that not only keeps pace with quantum hardware improvements but can preemptively gauge the quantum circuit processing costs of future device architectures. Being open-source, Benchpress ensures the transparency and verification of performance claims.
\end{abstract}
\date{\today}

\maketitle

\section*{Introduction}\label{sec:intro}

The promise of quantum computation lies in its ability to perform specific tasks more efficiently than otherwise possible using classical methods alone.  However, quantum computers do not operate in isolation and require classical computing resources for data pre- and post-processing.  As quantum computers continue to grow in size, it is imperative that the associated classical computing costs be evaluated for scalability, and the construction, manipulation, and optimization of quantum circuits play an out-sized role in terms of classical resource overhead.  Moreover, the performance of quantum computation software for output circuit quality, run-time, and memory consumption is critical to successful adoption of this technology. There is a variety of quantum software development kits (SDKs) that perform all or some fraction of quantum circuit pre-execution processing, such as Braket \cite{braket:2020}, BQSKit \cite{younis:2021}, Cirq \cite{cirq}, Qiskit \cite{javadi:2024}, Qiskit Transpiler Service (QTS) extension \cite{kremer:2024}, Staq \cite{amy:2020}, and Tket \cite{sivarajah:2021}, amongst others, that focus primarily on quantum circuit construction, manipulation, and optimization.  Here we define an SDK to be a collection of software development tools within a single installable package. The need to evaluate the performance of these software stacks has led to the creation of several collections of assessment circuits \cite{amy:2019,li:2023, mori:2023} and Hamiltonians \cite{sawaya:2023}, as well as some examples of how to execute collections of circuits in a benchmarking framework \cite{kharkov:2022, quetschlich:2023}.  

However, these earlier works have several limitations.  First, collections of quantum circuits are usually stored in OpenQASM \cite{cross:2022} compatible format and thus do not consider the performance of circuit synthesis \cite{yan:2024} in the transpilation process.  Addressing this need requires abstract circuits that must be written directly in the language of the SDK, making testing across several SDKs more challenging.  In addition, previous testing frameworks are not made to accommodate the testing of circuits at large qubit counts and run-times, nor are flexible enough to allow for testing over collections of two-qubit entangling gate topologies (coupling maps), quickly adding additional SDKs, or storing diverse sets of output metrics. 

To date, there has been no systematic study of the relative performance of mainstream quantum computing SDKs over extensive collections of quantum circuits at scale, nor has there been a flexible method by which these comparisons can be easily generated.  Such a study is critical as quantum devices and experiments push towards 100-qubits or more \cite{zhang:2023,kim:2023,hongye:2023,majumdar:2023,shtanko:2023,yasuda:2023,chen:2023,farrell:2024,pelofske:2023,baumer:2023,acharya:2024,bluvstein:2024,farrell:2024b,shinjo:2024,miessen:2024,robledo:2024,montanez:2024,cadavid:2024,alevras:2024,sachdeva:2024}, and the differences in software performance and scaling become pronounced.  An open-source test suite that not only evaluates multiple SDKs at these scales but also provides an execution framework that yields uniform testing over quantum software with disparate feature sets and capabilities would help researchers, developers, and end users alike ascertain the relative value of quantum computing software packages when targeting current and future quantum computing devices. With fault-tolerant computation out of reach in the near-term, this initial version of Benchpress is focused on test cases and metrics compatible with execution on noisy quantum processors.

To address these needs we have developed Benchpress, a unique open-source collection of tests explicitly designed to measure the performance of quantum computing software for quantum circuit creation and transformation.  Benchpress stands out for its common framework that allows testing across three key areas: quantum circuit construction, manipulation, and optimization. Although the intersection of functionality varies dramatically across the quantum computing software landscape, Benchpress utilizes notional collections of tests called ``workouts'' that allow the full test-suite to be executed across any quantum SDK with tests defaulting to being skipped if they are not explicitly overwritten with an SDK specific implementation.  Benchpress is thus able to quantify not only relative performance metrics amongst SDKs but also the breadth of functionality in a given software package.  

To supplement functionality missing in other SDKs, Benchpress uses Qiskit \cite{javadi:2024} throughout its infrastructure, particularly its compatibility with other SDKs and OpenQASM import and export capabilities. In addition, Qiskit allows for generating reference implementations for constructs such as abstract backend coupling maps that can, with minimal effort, be consumable by the other SDKs.  This yields a uniform, less error-prone testing environment that simplifies benchmarking.  Moreover, in making Benchpress open-source, we endeavor to create an open and transparent platform by which progress in quantum computing software can be faithfully evaluated.

In this work, we present results running the initial version of Benchpress, evaluating $1066$ tests for each of seven different quantum SDKs considered. We give a detailed analysis of these results, test selection criteria, and highlight the key takeaways from the findings. Section.~(\ref{sec:methods}) explores the goals and methodology behind the testing framework and SDK-specific considerations.  Data and code availability are given in Sec.~(\ref{sec:data}) and (\ref{sec:code}), respectively.

\section{Results}\label{sec:results}

\begin{table}[b] \centering
\ra{1.3}
 \begin{tabular}{||r c||} 
 \hline
 SDK & Version number  \\ [0.5ex] 
 \hline\hline
 \texttt{amazon-braket-sdk} (braket) & 1.86.1  \\ 
 \texttt{bqskit} & 1.1.2 \\
 \texttt{cirq} & 1.4.1 \\
 \texttt{pytket} (tket) & 1.31.0\\
 \texttt{qiskit} & 1.2.0 \\
  \texttt{qiskit\_transpiler\_service}\footnote{After testing was completed this package changed names to \texttt{qiskit\_ibm\_transpiler}} & 0.4.8 \\
 \texttt{staq} & 3.5\\ [1ex] 
 \hline
 \end{tabular}
\caption{Software development kits (SDK) and corresponding version numbers used in generating the reported sample results.  Each version represents the latest release with pertinent functionality as of August 30, 2024 \footnote{Tket version $1.31.1$ includes only cosmetic changes to documentation.}.}
 \label{tbl:sdk}
\end{table}

This section presents results covering the seven SDKs listed in Table~(\ref{tbl:sdk}). Other packages, such as CUDA-Q \cite{cudaq}, are not included in this study as their feature sets at the time of testing were insufficient to accommodate the test cases; basic functionality such as computing the depth and number of operations of a circuit is not available without executing the circuit, it is unclear how to perform many circuit manipulations, and there is no predefined general-purpose compilation workflow at present. Benchpress is designed to be modular, and other SDKs beyond those considered here can be added straightforwardly.  Unlike the other SDKs, the QTS augments the transpilation tools of Qiskit with reinforcement learning methods for Clifford synthesis and qubit routing, and thus is better viewed as an extension to Qiskit rather than a competing package.

\begin{figure}[t]
\centering
\includegraphics[width=8cm]{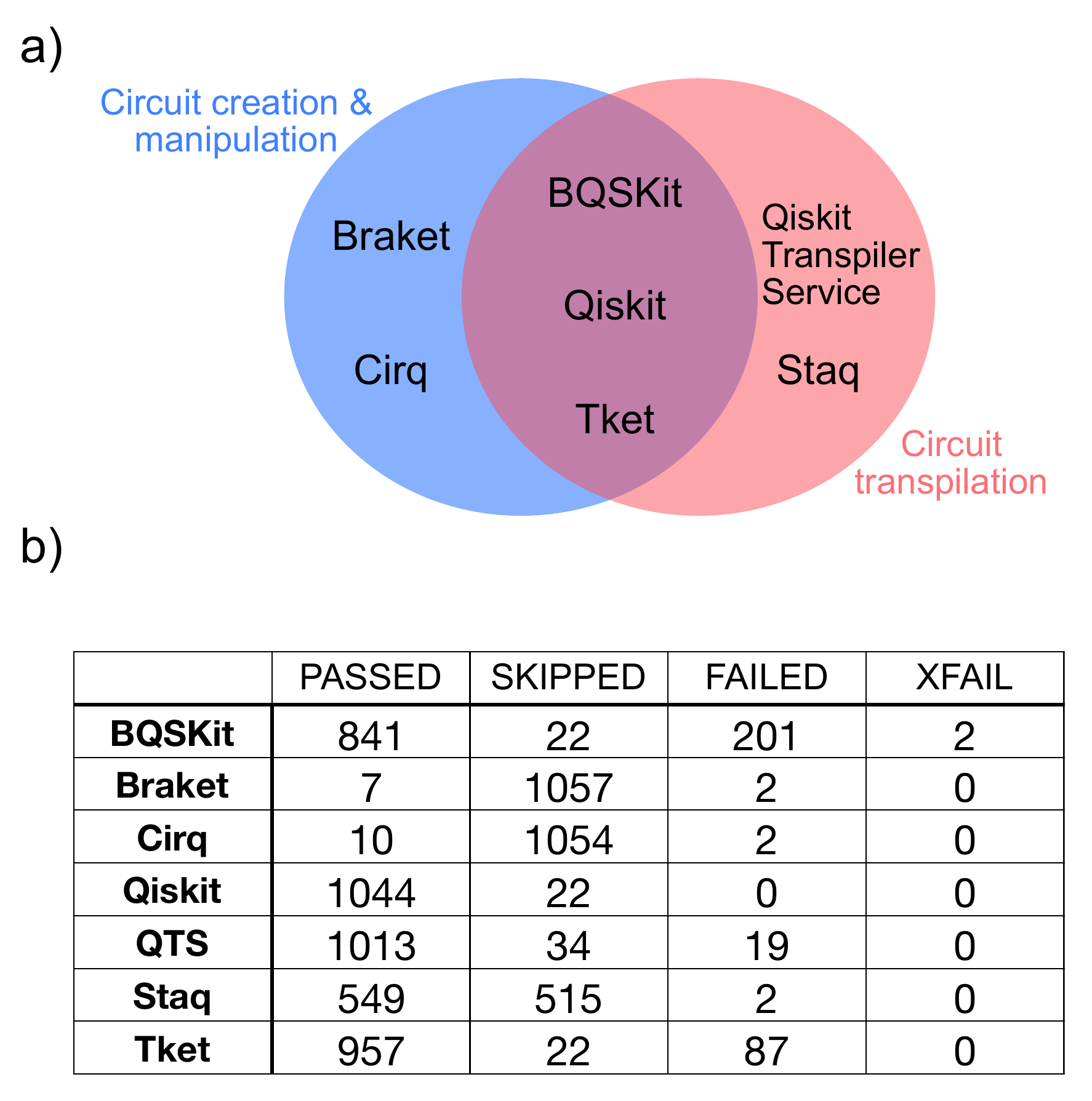}
\caption{a) Venn diagram of SDK functionality. b) Status of all $1066$ tests for each SDK based on the definitions given in Sec.~(\ref{sec:results}).  $22$ tests are universally skipped due to insufficient qubit count for the target used in device transpilation [Sec.~(\ref{sec:trans})].}
\label{fig:venn}
\end{figure}

The breadth of functionality available in the tested SDKs varies widely and must be accounted for when benchmarking. In general, current quantum software packages can be categorized as having the ability to create and manipulate circuits and/or offering predefined transpilation toolchains that map quantum circuits to quantum hardware systems. This Venn diagram of functionality is shown in Fig.~(\ref{fig:venn}a).  

Benchpress accommodates SDKs with disparate feature sets by running the full test suite over each SDK, regardless of whether the individual tests are supported. Our test environment is based on \texttt{pytest} \cite{pytest}, and we map each of the standard  \texttt{pytest} output types to the following definitions:

\begin{itemize}
    \item \textbf{PASSED} -  Indicates that the SDK has the functionality required to run the test, and doing so completed without error and within the desired time limit, if any.
    \item \textbf{SKIPPED} - The SDK does not have the required functionality to execute the test, or the test does not satisfy the problem's constraints, e.g., the input circuit is wider than the target topology. This is the default status for all notional tests.
    \item \textbf{FAILED} - The SDK has the necessary functionality, but the test failed or was not completed within the set time limit, if any.
    \item \textbf{XFAIL} - The test fails irrecoverably. It is therefore tagged as ``expected fail" rather than being executed \footnote{Because we execute tests in subprocesses to implement a timeout mechanism, some failures can kill the subprocess but otherwise not affect the remaining tests. These are considered \texttt{FAILED} per the definition here.}. For example, a test is trying to use more memory than available.
\end{itemize}

All tests have notional definitions called ``workouts", see Sec.~(\ref{sec:methods}) that are placeholders for SDK-specific implementations and default to  \texttt{SKIPPED} unless explicitly overwritten in each SDK test suite.  In this way, Benchpress can use skipped tests as a proxy for measuring the breadth of SDK functionality, and this can be tracked automatically when additional functionality is added in the form of new tests.  Figure~(\ref{fig:venn}b) shows the distribution of tests by status for the results presented here when running the full Benchpress test suite against each SDK.  While the specifics of test failures will be discussed below, we note that $97\%$ of failures occur when running benchmarks originating from other test suites, as opposed to those created explicitly for Benchpress. 

In what follows, we break the test suite into two components: grouping circuit creation and manipulation and evaluating circuit transpilation tests as a whole. All results presented here are generated using an AMD $7900$ processor with $128~\rm{Gb}$ memory running Linux Mint $21.3$.  SDKs were run on Python $3.12$ with the version numbers shown in Table~(\ref{tbl:sdk}). Complete system information is recorded in the JSON files output by Benchpress and is available at the location given in Sec.~(\ref{sec:data}).

\subsection{Circuit construction and manipulation}\label{sec:construct}

\begin{figure}[t]
\centering
\includegraphics[width=7.75cm]{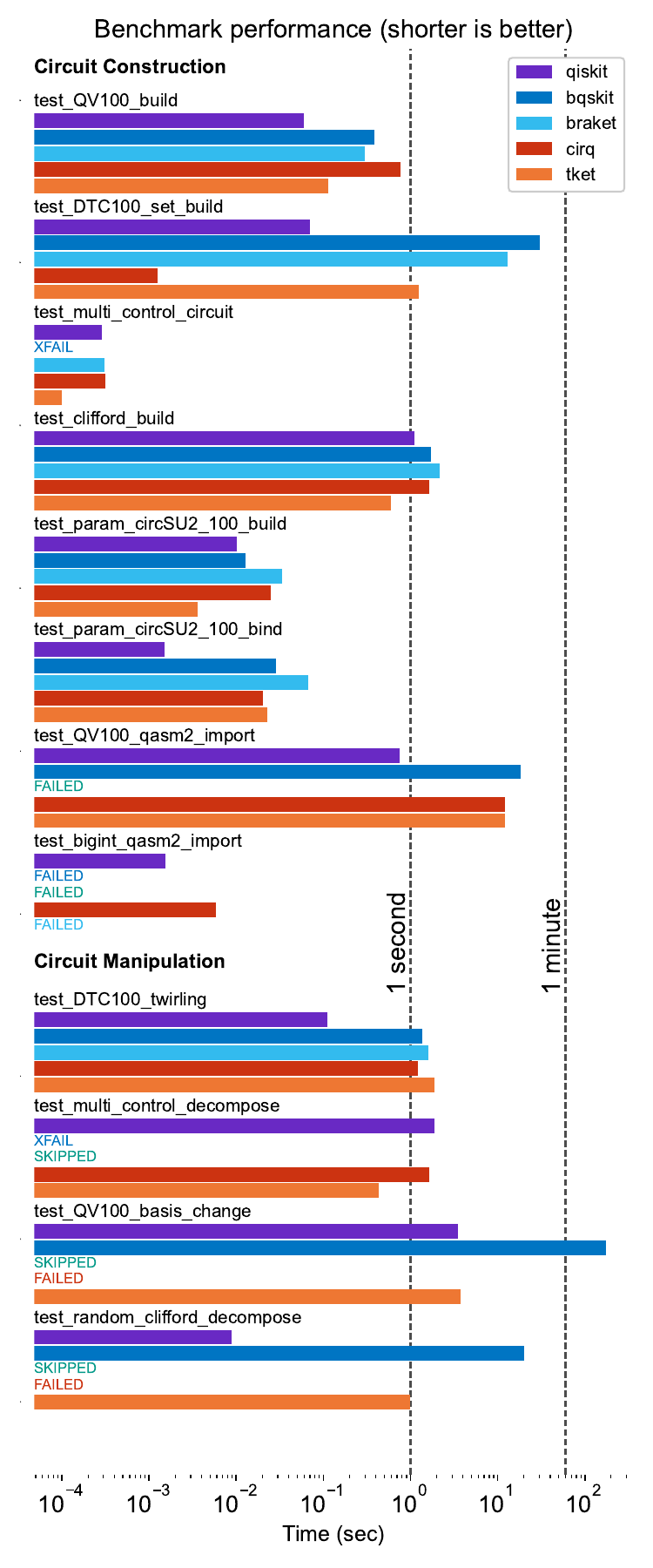}
\caption{Benchmark results for the circuit construction and manipulation portions of Benchpress.  Tests that are \texttt{SKIPPED} or marked as \texttt{FAILED} or \texttt{XFAIL} are labelled accordingly.}
\label{fig:construct}
\end{figure}

Although nominally a tiny part of an overall circuit processing workflow time budget, as compared to circuit transpilation, measuring the timing of circuit construction and manipulation gives a holistic view of quantum SDK performance.  Moreover, if suitably chosen, such tests can provide insights into other parts of the entire circuit compilation process.  The present version of Benchpress includes 12 such tests, with tests aimed at representing scenarios encountered during real-world SDK usage.  Circuit construction includes eight tests that look at timing information needed to build 100-qubit circuits for families of circuits such as Quantum Volume (QV) \cite{cross:2019} (\texttt{test\_QV100\_build}), Hamiltonian simulation \cite{zhang:2023} (\texttt{test\_DTC100\_set\_build}), random Clifford circuits \cite{BM2021} (\texttt{test\_clifford\_build}), 16-qubit iterative construction of multi-controlled gates (\texttt{test\_multi\_control\_circuit}), and parameterized ansatz circuits with circular entangling topology (\texttt{test\_param\_circSU2\_100\_build}).  We will reuse many of these circuits in device benchmarking.  We also include the time to bind values to parameterized circuits (\texttt{test\_param\_circSU2\_100\_bind}) and import from OpenQASM files into the construction category.  This latter set of QASM tests includes importing a 100-qubit QV circuit and reading a file with an integer corresponding to a 301-bit classical register, \texttt{test\_QV100\_qasm2\_import} and \texttt{test\_bigint\_qasm2\_import} respectively.  Our focus on $100$-qubit circuits stems from the need for sufficient complexity for gathering faithful timing information, and the fact that these circuits are within the number of qubits available on present-day quantum processors.

Circuit manipulation is the set of operations that can be performed on a fully built circuit. Out of the four such tests included, two represent basis transformations, \texttt{test\_QV100\_basis\_change} and \texttt{test\_random\_clifford\_decompose} \cite{Bravyi2021cliffordcircuit}, taking an input OpenQASM file and expressing them in a differing set of gates.  In a similar vein, we use the same multi-controlled circuit used in the circuit construction tests and time the decomposition into a QASM-compatible gate set in \texttt{test\_multi\_control\_decompose}. In contrast to the previous tests, this decomposition requires a non-trivial synthesis step and provides additional insight into how well the SDKs transform abstract quantum circuits into primitive components.  This is captured in the number of two-qubit gates in the circuit returned at the end of the test, and this value is also recorded.  Finally, we also implement Pauli-twirling \cite{knill:2004, wallman:2016} in each SDK, \texttt{test\_DTC100\_twirling}, recording the time it takes to twirl $19800$ CNOT gates in a Hamiltonian simulation circuit.

While there is no clear winner when looking at each test in isolation, the aggregate timing information and number of failed or skipped tests tell a different story.  Qiskit is the only SDK that passes all of the circuit construction tests, and does so in a time of $2.0$ seconds.  The next closet competitor is Tket, which completes all but one test in $14.2$ seconds.  BQSKit fails two tests, and clocks the slowest total completion time over passed tests at $50.9$ seconds.  Results worth highlighting include Cirq's performance at constructing a set of Hamiltonian simulation circuits (\texttt{test\_DTC100\_set\_build}) in a time $55\rm{x}$ faster than the nearest competitor Qiskit.  Likewise, Qiskit outperforms the other SDKs in the parameter binding test (\texttt{test\_param\_circSU2\_100\_bind}), recording a time $13.5\rm{x}$ faster than the next closest SDK.

The most notable feature in Fig.~(\ref{fig:construct}) is the number of skipped, fails, or expected-fail tests.  Here, we address those issues.  Starting at the top of Fig.~(\ref{fig:construct}), the first \texttt{XFAIL} test is for BQSKit on multi-controlled circuit building.  BQSKit heavily uses dense numerical linear algebra throughout its compilation pipeline, and the 16-qubit multi-controlled X-gate used in the test took more memory than the test machine had ($128\rm{Gb}$).  A similar reason is behind the other \texttt{XFAIL} for BQSKit in the multi-controlled decomposition manipulation test (\texttt{test\_multi\_control\_decompose}).  Next, Braket failed the QV OpenQASM import test because there is no native support for the \texttt{qelib1.inc} file that is a standard include file in OpenQASM 2, e.g., both the Feynman \cite{amy:2019} and QasmBench \cite{li:2023} collections of circuits use this include file.  Finally, 3 of 5 SDKs could not load an OpenQASM file with an integer larger than 64 bits.  Here, BQSKit failed because of the lack of dynamic circuit support, Braket failed in the same manner as the previous QASM test, and the C++ JSON library used in Tket lacks support for these "BigInt" numbers.  

For circuit manipulation, both Qiskit and Tket complete all tests, with Qiskit completing all four tests in $5.5$ seconds versus $7.1$ for Tket.  The large number of skipped tests for Braket is due to the lack of basis transformation capabilities. However, add-on packages allow this to be done in other SDKs and then passed back to Braket \cite{braketprovider}.  The two failed tests for Cirq are due to using the basis set of gates: $RZ$, $X$, $\sqrt{X}$, and $CZ$, which despite being supported gates, hits a recursion limit. In addition to timing information, the multi-controlled decomposition test allows for examining the quality of the synthesis algorithms in each SDK via the number of two-qubit gates in the output circuit.  Here, Tket produced the circuit with the fewest two-qubit gates, $4457$, versus $7349$ and $17414$ for Qiskit and Cirq, respectively.  The relative gate counts and depth for the circuit manipulation tests are shown in Supplementary Fig.~(\ref{supp-fig:stats}).

\begin{figure*}[t]
\centering
\includegraphics[width=\textwidth]{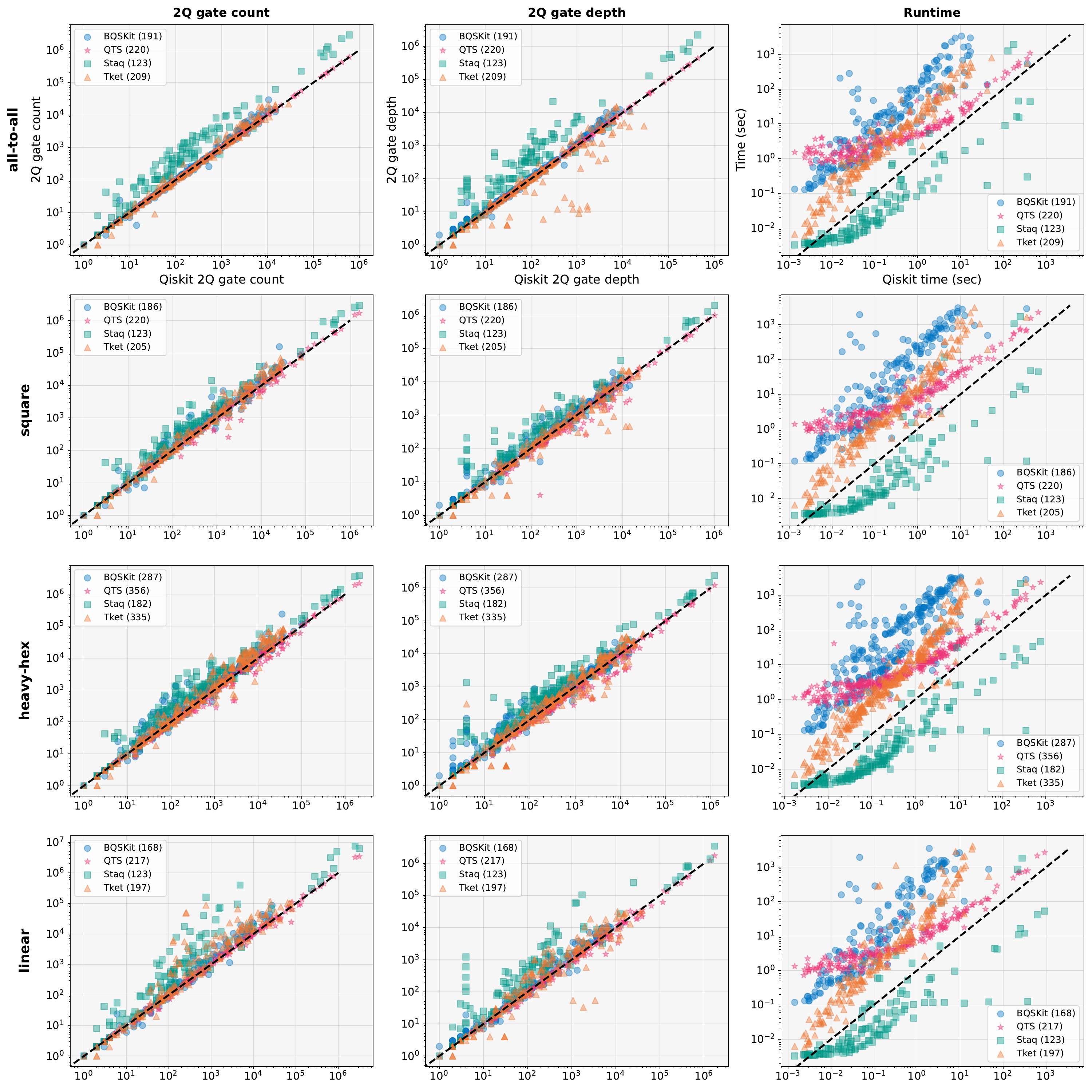}
\caption{Results generated from the device- and abstract transpilation tests in Benchpress.  Rows indicate the topology used in the tests, whereas columns label the reported metric.  For brevity, only the top row of plots is labeled.  The target topology for the device transpilation tests is ``heavy-hex".  Markers above the dashed line indicate tests where Qiskit performs better on a given metric relative to a the specified SDK.  In contrast, markers below highlight SDK performance that is better than that of Qiskit.  The total number of passed tests in each data set is given in the legends.}
\label{fig:results}
\end{figure*}

\begin{table*}[t]
\centering
\scriptsize
\ra{1.3}
\begin{tabular}{@{}rccccccccccccccc@{}}\toprule
& \multicolumn{4}{c}{\textbf{BQSKit}} &  \multicolumn{4}{c}{\textbf{QTS}}  & \multicolumn{4}{c}{\textbf{Staq}}  & \multicolumn{3}{c}{\textbf{Tket}}\\
\cmidrule{2-4} \cmidrule{6-8} \cmidrule{10-12} \cmidrule{14-16}

& ~2Q gates~ & ~2Q depth~ & ~time~ && ~2Q gates~ & ~2Q depth~ & ~time~ && ~2Q gates~ & ~2Q depth~ & ~time~ && ~2Q gates~ & ~2Q depth~ & ~time~\\ \midrule

\textbf{All tests*} & $1.26$/$1.13$ & $1.27$/$1.18$ & $108$/$98.0$ && $0.96$/$1.0$ & $0.89$/$1.0$ & $18.7$/$14.5$ && $2.8$/$2.45$ & $2.91$/$2.40$ & $0.26$/$0.22$ && $1.31$/$1.09$ & $0.98$/$1.00$ & $13.3$/$12.7$ \\

\textbf{all-to-all} & $1.04$/$1.00$ & $1.11$/$1.05$ & $75.0$/$71.0$ && $1.0$/$1.0$ & $1.0$/$1.0$ & $18.3$/$15.8$ && $3.35$/$3.76$ & $4.02$/$4.07$ & $0.36$/$0.34$ && $1.08$/$1.00$ & $0.75$/$1.00$ & $14.2$/$14.5$ \\

\textbf{square} & $1.31$/$1.25$ & $1.30$/$1.22$ & $104$/$112$ && $0.96$/$1.0$ & $0.85$/$0.97$ & $19.1$/$14.8$ && $2.15$/$2.13$ & $2.36$/$2.14$ & $0.23$/$0.19$ && $1.21$/$1.12$ & $0.95$/$1.00$ & $12.4$/$12.1$ \\

\textbf{heavy-hex} & $1.50$/$1.32$ & $1.47$/$1.29$ & $125$/$129$ && $0.90$/$1.0$ & $0.82$/$0.92$ & $18.6$/$13.5$ && $2.27$/$2.09$ & $2.31$/$1.77$ & $0.21$/$0.19$ && $1.30$/$1.16$ & $1.04$/$1.07$ & $12.0$/$11.2$ \\

\textbf{linear} & $1.23$/$1.08$ & $1.25$/$1.20$ & $115$/$92.1$ && $0.98$/$1.0$ & $0.91$/$0.99$ & $21.5$/$17.0$ && $3.92$/$3.48$ & $3.66$/$2.75$ & $0.30$/$0.31$ && $1.76$/$1.31$ & $1.21$/$1.10$ & $17.2$/$14.5$ \\

\bottomrule
\end{tabular}
\caption{Geometric mean \cite{stats} / median values for SDK performance metrics, normalized to the corresponding Qiskit values.  Results are presented over all-tests combined, as well as sorted by target device topology. (*) Here, ``all-tests" for BQSKit, QTS, and Tket includes all non-failing tests out of a total possible $1032$.  In contrast, Staq results are over the subset of $551$ tests that do not include a synthesis step.}
\label{tbl:results}
\end{table*}

\subsection{Device and abstract transpilation}\label{sec:trans}

Due to their vast array of possible input parameters and a large fraction of overall runtime, transpilation tests form the bulk of the tests in Benchpress. We split these tests into two groups depending on whether they target a model of a real quantum device, or if their target is an abstract topology defined by a generating function.  We label these as ``device" and ``abstract" transpilation tests, respectively.  These tests differ because device testing targets a fixed model of a quantum system, regardless of input quantum circuit size, and includes error rates that can be utilized in noise-aware compilation routines.  Noise aware heuristics can have an impact on both the duration of the compilation process, as well as the two-qubit gate count and depth; they can paradoxically lead to worsened performance if applied overly aggressively.  However, because we do not execute the resulting circuits on hardware, the impact of these techniques on output fidelity is not included. In contrast, abstract transpilation tests take an input circuit and finds the smallest topology compatible with the circuit.  In this manner, we can benchmark SDKs across arbitrary circuit sizes and topology families, and allowing for user configuration of the basis gates in the \texttt{default.conf} file, see Sec.~(\ref{sec:methods}).

Device transpilation tests come from three sources.  First, we include a collection of tests that primarily focus on $100$-qubit circuits representing circuit families such as QV, Quantum Fourier Transform (QFT), Bernstein-Vazirani (BV), and random Clifford circuits.  In addition, circuits generated from Heisenberg Hamiltonians over a square lattice and Quantum Approximate Optimization Algorithm (QAOA) circuits corresponding to random instances of a Barabási-Albert graph are also included.  We also add $100$- and $89$-qubit instances of the same parameterized ansatz circuits used in Sec.~(\ref{sec:construct}), where the former can be embedded precisely on a heavy-hex device, i.e., there is an ideal answer, while the latter cannot.  This set also includes a circuit with a BV-like structure, but where the circuit can be optimized down to single-qubit gates if transpiled appropriately.  Because this set of circuits is represented in OpenQASM form or generated using QASM-compatible gates only, they do not test the synthesis properties of each SDK.  To do so, we include a set of $100$ abstract circuits using Hamiltonians included in the HamLib library \cite{sawaya:2023} for time evolution.  The choice of Hamiltonians is described in Sec.~(\ref{sec:ham}), and results in a set of Hamiltonians from two to $930$ qubits in size.  Finally, we include the Feynman collection \cite{amy:2019} of circuits that are up to $768$ qubits, and also OpenQASM-based, in device transpilation tests.  Depending on the target quantum system for device transpilation, some device tests may be skipped due to insufficient physical qubit count. For benchmarking against abstract topologies, we run the same set of Hamiltonian simulation circuits run for device transpilation and include OpenQASM tests from QasmBench \cite{li:2023} that go up to $433$ qubits.  
 
Our performance metrics for both sets of tests are two-qubit ($2$Q) gate count, $2$Q-gate depth, and transpilation runtime.  In addition, we record the number of qubits in the input circuit, QASM load time (if any), and the number and type of circuit operations at the output.  Any additional metrics that are compatible with JSON serialization can be included.  The target system used in the device transpilation tests is the \texttt{FakeTorino} system, which is a snapshot of a 133-qubit Heron system from IBM Quantum that includes calibration data suitable for noise-aware compilation.  Abstract topologies tested are ``all-to-all", ``square", ``heavy-hex", and ``linear", which includes most typical device topologies, and are predefined graphs in the \texttt{rustworkx} library \cite{rustworkx}.  We have configured the abstract models to use the basis set \texttt{[`sx', `x', `rz', `cz']}.  Finally, to limit the duration of the tests, we have set a timeout limit of $3600$ seconds (one hour), after which the test is marked as \texttt{FAILED}.

In this work we focus on testing the predefined transpilation pipelines in each SDK.  In this way, we aim to measure the relative performance that a typical user would see, and eliminate the bias involved when creating bespoke transpilation workflows in SDKs of which we have less knowledge than Qiskit. In making Benchpress open-source, we hope comparisons of optimal performance can be lead by community experts in each SDK.  Here we use the default optimization values for both Tket ($2$) and BQSKit ($1$).  Qiskit does not have a well defined default optimization level, with the \texttt{transpile} function having a default value of $1$, whereas the newer \texttt{generate\_preset\_passmanager} interface must have the optimization level explicitly set.  In this work, we use optimization level $2$ for Qiskit that will be the default value for both ways of calling the transpiler functionality starting in version \texttt{1.3.0}. This same optimization value is used for the QTS as well.  As discussed in Sec.~(\ref{sec:staq}), Staq was set to optimization level $2$ in order to generate circuits valid for the target topologies.  All other transpiler values are left unchanged.

Figure~\ref{fig:results} shows the results for the five SDKs that support transpilation, see Fig.~(\ref{fig:venn}), over the combined set of device and abstract tests.  Out of $1054$ total tests, $22$ are device transpilation tests larger than the target Heron device and are \texttt{SKIPPED} regardless of the SDK.  In addition, the Staq compiler takes OpenQASM files as input and thus is unable to execute the Hamiltonian simulation tests, defaulting to those tests being skipped.  Only passing tests are shown in Fig.~(\ref{fig:results}).  As Qiskit is the only SDK that passes the entire set of transpilation tests, we use those results as a baseline when presenting results.  In what follows, we focus on the statistics of ensembles of circuits as a whole, as opposed to looking at individual tests cases.  A lower-level analysis can be done using the published results [Sec.~(\ref{sec:data})].  A break down of the test results for each of the open-source test libraries used here is presented in the Supplemental Material Sec.~(\ref{supp-sec:groupings}).

Figure~\ref{fig:results} breaks the results apart by target topology for each of the three reported metrics. The dashed diagonal line represents the Qiskit baseline above which any data points indicate that the specified SDK test result is worse than the corresponding Qiskit value. On the other hand, those markers below this line highlight an SDK performing better than Qiskit.   Table~\ref{tbl:results} numerically quantifies these results, tabulating the statistical values of results across each SDK and topology. 

From Fig.~(\ref{fig:results}) and Tbl.~(\ref{tbl:results}), we see a few trends emerge.  First, we see that, as a whole, Qiskit outperforms BQSKit, Staq, and Tket in terms of 2Q gate count; a trend clearly seen in the BQSKit and Tket data. This is particularly true when targeting topologies with limited connectivity.  In contrast, while the two-qubit gate depth is worse than Qiskit for BQSKit and Staq, Tket out-performs Qiskit overall, with prominent gains for more connected topologies.  In particular, there is a collection of Hamiltonian simulation tests on which Tket yields substantial 2Q depth reduction relative to Qiskit.  This set is the largest for ``all-to-all" coupling topologies, suggesting that this improvement comes from the synthesis steps performed by Tket.  This is corroborated by the synthesis results in Sec.~(\ref{sec:construct}).  Data for other topologies indicate that this improved synthesis becomes less critical as the connectivity decreases and circuit routing becomes dominant.  Note that for transpilation pipelines that include stochastic components, such as the Sabre-based routing routine in Qiskit, both the number and depth of two-qubit gates can vary across different transpilations of the same circuit.  When looking at a single test case, this randomness can have a large impact on relative performance numbers.  However when looking across large collections of tests, as done here, the fluctuations between runs approximately average out.  We have confirmed that this is the case by running multiple instances of the test suite for Qiskit and found that the results presented here do not change appreciably between runs. 

The QTS utilizes much of the Qiskit transpilation pipeline internally, differing only in the use of a routing method based on reinforcement learning \cite{kremer:2024}.  Therefore, we expect to see identical results for 2Q gate count and depth between Qiskit and the QTS for the "all-to-all" coupling tests.  This is confirmed in  Tbl.~(\ref{tbl:results}).  In terms of 2Q-gate count, the QTS performs a few percent better than Qiskit on restricted coupling maps, with on average an $11\%$ reduction on heavy-hex topologies.  However, larger gains from the QTS appear in the 2Q-gate depth where a geometric mean improvement of $12\%$ over Qiskit is observed over all topologies, and up to $22\%$ when looking solely at heavy-hex results.  These results are inline with the fact that the QTS routing step is trained only over heavy-hex topologies, and thus is expected to perform better there.  Like the Tket 2Q-depth results, the performance gains from the QTS come from a handful of tests with markedly lower depths than the corresponding Qiskit circuits, as opposed to an across the board advantage.  However, in contrast to Tket where the synthesis step is responsible for the improved depth characteristics, the QTS benefits are isolated to the routing stage.

The differences in test runtimes shown in Fig.~(\ref{fig:results}) clearly show that Staq is the fastest of the transpilation pipelines, while the QTS and Tket are over an order of magnitude slower that Qiskit.  BQSKit has the longest runtimes, performing two orders slower than Qiskit.  Note that for circuits over restricted topologies that take $\sim 10~\rm{sec}$ or more, as measured by Qiskit, the corresponding Tket time begins to diverge, suggesting an unfavorable scaling for the routing algorithm used by Tket compared to that of Qiskit.  The results for Staq are an excellent example of the quality-vs-time trade-off made when designing transpilation pipelines; it is possible to gain performance at the cost of reduced circuit quality or vice versa.  As discussed in Sec.~(\ref{sec:staq}), Staq also does not perform the same collection of steps as the other SDKs.  Being a service, the QTS has a minimum input-output (IO) time of a few seconds, as seen in Fig.~(\ref{fig:results}) for tests with short Qiskit runtimes.  In contrast, this IO overhead is immaterial for more complex circuits that take longer time, and it is seen that the QTS runtime remains longer than that of Qiskit irrespective of the target topology, but this overhead is approximately constant for each topology.

While Qiskit completes the full set of transpilation tests in $0.18$ days, the failed tests for the other SDKs makes a direct timing comparison impractical.  However, using the relative timing information in Tbl.~(\ref{tbl:results}) together with the corresponding Qiskit test times, we can estimate the full transpilation test suite time of the remaining SDKs. As the fastest SDK tested, the estimated full time of the Staq tests comes in less than Qiskit at $0.15$ days.  In contrast, BQSKit, QTS, and Tket are estimated to run more than a day at $20.5$, $1.3$, and $4.73$ days, respectively.

Because we target a fake IBM Quantum system that includes timing information, we can use the device transpilation tests to gauge the relative timing for circuit transpilation verses that of actual execution on hardware.  To this end, we schedule each Qiskit circuit to obtain the execution time on chip, and add the idle time between circuits, called the \texttt{default\_rep\_delay} in Qiskit, to get the total time per circuit execution \footnote{This data was added after the initial version of Benchpress, and is not included in the published results.}.  Multiplying this value by the default number of executions for IBM Quantum hardware, currently $4096$, gives us a good estimate for circuit execution time without requiring hardware usage.  For the Qiskit device transpilation tests, the total compilation time was $1055$~sec, whereas the total execution time of all circuits would be $968$~sec.  Restricting ourselves to circuits with $\geq 100$ qubits the total compilation time is $\sim 1.4x$ longer than the $78$~sec hardware runtime; the compilation costs begin to dominate at larger qubit counts. Table~(\ref{tbl:results}) shows that other comparable SDKs such as BQSKit and Tket take one- to two-orders of magnitude longer to compile circuits, while only marginally increasing the overall circuit depth, indicating that the ratio of compilation to run-time is more pronounced when using these compilation stacks to target the same quantum device.  Note however that is conclusion is dependent on the targeted quantum hardware modality.  For example, platforms such as trapped-ion systems have greater connectivity and, in general, require less compilation time because of this.  In addition, trapped-ion run-times are two-orders of magnitude or more longer than those on superconducting devices, e.g. see Ref.~(\cite{lotstedt:2024}), and therefore the ratio of compilation to hardware run-times may be less than those presented here when targeting diffing hardware platforms.
 
Not included in Fig.~(\ref{fig:results}) and Table~(\ref{tbl:results}) is information on skipped or failed tests.  Outside of the $22$ device transpilation tests that do not fit the target device, Qiskit is the only SDK that passed the full $1032$ collection of tests.  BQSKit failed $200$ (19\%), QTS $19$ (2\%), Staq $2$ (0.3\%), and Tket $86$ (8\%) tests.  Note that, as an OpenQASM-based compiler, Staq can only execute the $551$ tests that do not include synthesis.  These failures are, in large part, due to the timeout limit set on the tests.  However, additional failure modes include QASM parsing issues in BQSKit, IO service errors in the QTS, C++ errors in Tket, and circuit validation failures in Staq, where the output circuit did not match the topology of the target device.

\subsection*{Hamiltonian selection criteria}\label{sec:ham}
Hamiltonians included in Benchpress originate from the HamLib Hamiltonian library  \cite{sawaya:2023}, which includes problems from chemistry, condensed matter physics, discrete optimization and binary optimization.  We randomly selected Hamiltonians from HamLib to be included in the benchmark suite presented in this work. The random selection is biased towards reflecting the distribution of Hamiltonian characteristics prevalent in HamLib such as number of qubits and number of Pauli terms. Furthermore, we limited the number of qubits in the selected Hamiltonians to $\le 1092$ and the number of Pauli terms to $10,000$ or fewer. Furthermore, the random selection is biased towards ``unique" Hamiltonians; the selection of different encodings of the same Hamiltonian is discouraged. One hundred HamLib Hamiltonians are included in this version of Benchpress, where 35 Hamiltonians are from chemistry and condensed matter physics problem classes each, and 15 Hamiltonians are chosen from both discrete and binary optimization problem classes.

\section{Discussion}\label{sec:discussion}
We have presented Benchpress, an open-source execution framework and collection of tests for evaluating the performance of quantum computing software in a unified manner.  Benchpress allows for accommodating a wide variety of quantum computing frameworks, leveraging the interoperability of Qiskit to provide a uniform testing environment, irrespective of SDK functionality. Using an initial data set of $1066$ tests, we have determined the performance of seven quantum SDKs over a wide range of circuit families, qubit counts, and device topologies for common performance metrics, and highlighted the salient features of the resultant datasets of circuits.  Including the metrics considered here, Benchpress also allows for measuring the memory consumption of SDKs for each test case using Memray \cite{memray}.  However, this analysis comes with a substantial overhead, exacerbating runtimes and test suite memory consumption.  As such, this functionality is only useful when looking at small targeted sets of test cases.

Benchpress is designed for transparent and reproducible comparisons between quantum SDKs.  Benchpress, along with all of the results presented here, are open-sourced, and can be readily executed by anyone looking to validate our findings.  Indeed, our aim is to make the benchmarking process community-driven, with experts fluent in each SDK helping to refine the testing process or push it in new directions.  Unlike most other benchmarking frameworks, we envision a fluid approach to quantum SDK testing, with tests being refined as quantum hardware and software mature, and target performance metrics being adjusted or expanded as necessary, using Benchpress version numbering to track the underlying test suite changes.

In the end, this work aims to benefit the community of quantum researchers, developers, and end-users at large by providing a trusted source by which absolute and/or relative improvements in quantum software can be faithfully examined.  For researchers and developers this can mean highlighting performance bottlenecks that need to be improved.  For example, the small number of circuits on which both Tket and QTS outperform Qiskit in terms of 2Q-gate depth are of interest to us, and identify areas for improved circuit synthesis and routing, respectively in Qiskit.  For others, this work may serve as a guide for where to target future SDK improvements for improving runtime and/or fixing coding bugs. As an open-source project, Benchpress is aimed at benefiting the quantum community as a whole, and we welcome contributions in the form of bug fixes, code improvements, and new tests cases at the projects Github website \cite{benchpress}. Finally, end-users can use this work as a guide for selecting the appropriate SDK, or perhaps combinations of SDKs, that are optimal for a given task.  In all cases, shedding light on the performance of quantum computing software can only push the field further and aid in the successful adoption of this computing paradigm.

\section{Methods}\label{sec:methods}

\begin{figure*}[t]
\centering
\includegraphics[width=15cm]{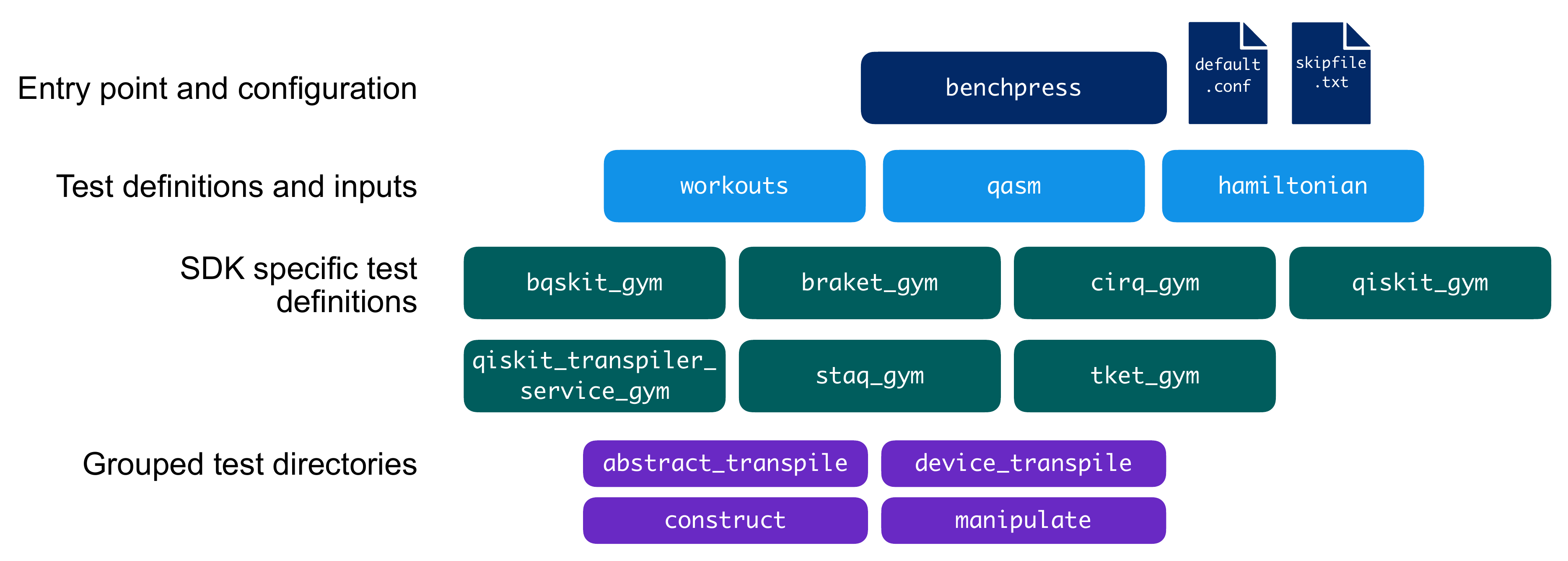}
\caption{Organization of the Benchpress data set. The entry point is the \texttt{benchpress} directory, and SDK specific flags and options are set in the \texttt{default.conf} file. Optionally, tests taking longer than a specified timeout can be automatically included in the \texttt{skipfile.txt}. Abstract test definitions are included in the \texttt{workouts} directory, and test inputs in the form of OpenQASM files or Hamiltonians are included in the \texttt{qasm} and \texttt{hamiltonian} folders, respectively. Tests specific to each SDK are located in the corresponding \texttt{*\_gym} directories. Inside each ``gym", tests are organized in groups based on the target functionality to be tested.}
\label{fig:org}
\end{figure*}

The organization of Benchpress is shown in Fig.~(\ref{fig:org}). The test suite does not need to be installed, but executing the tests requires \texttt{pytest} and any SDK-specific dependencies. In addition, we use \texttt{pytest-benchmark} \cite{benchmark}, which provides a robust method for timing tests and generating result information in JSON format. However, because these tools are typically used in the context of unit testing, they are primarily aimed at tests requiring only short durations of time. Here, we are interested in the opposite regime where test times can easily approach an hour or more and, in some cases, have run for a week before being manually terminated. With the large number of tests considered here, running all tests to completion is impractical. As such, we run all tests using a modified version of \texttt{pytest-benchmark} that wraps each test in a subprocess that can optionally be terminated after a specified time out. Tests that exceed a time-out are automatically added to the \texttt{skipfile.txt} shown in Fig.~(\ref{fig:org}). This file dramatically reduces the overhead from repeated test evaluations. However, it is tied to the specific computer on which it is generated and can mask performance improvements if used across different SDK versions. The file used in this work was generated using the same SDK versions given in Table~(\ref{tbl:sdk}). Note that using subprocesses to enforce timing can have adverse effects when timing software uses parallel processing. As such, the modified version of \texttt{pytest-benchmark} used for each SDK differs in the mechanism by which it spawns processes.

In order to define a uniform collection of tests across SDKs, Benchpress uses abstract classes of tests called ``workouts", where each test is defined as a method to a Python class, where the name of the method defines the test name, and each test is decorated with \texttt{@pytest.mark.skip} by default. Each class of tests is also logically organized into groups using \texttt{pytest-benchmark}. Implementing the actual tests for a given SDK requires overloading the abstract definitions in the workouts with a specific implementation. These are included in the "gym" directory corresponding to the given SDK, see Fig.~(\ref{fig:org}), and grouped into their respective categories. We have included a verification mechanism that verifies that only those tests defined in the workouts are allowed to be present at the gym level. We enforce this gym partitioning so that each SDK can be run in isolation, and we execute the tests in each gym in a separate environment. However, to make a uniform testing experience across SDKs, Benchpress makes use of Qiskit throughout its infrastructure, in particular its compatibility with other SDKs and OpenQASM import and export capabilities, to supplement functionality missing in other SDKs, as well as provide reference implementations for data like abstract backend entangling gate topologies that can, with minimal effort, be consumable by the other SDKs. Thus, Qiskit is a requirement common across all SDKs, but the version of Qiskit can differ, needing to satisfy the minimal requirements only.

Customizing the execution process is done via a configuration file, \texttt{default.conf}, that allows for setting Benchpress-specific options such as the target system used for device transpilation, as well as the basis gates and set of topologies utilized for the abstract transpilation tests, see Sec.~(\ref{sec:trans}). We have decided to allow multiple device topologies within a given benchmark run, but the basis gates are fixed throughout. This choice is motivated by the fact that we explicitly focus only on the number and depth of two-qubit gates in a circuit. With most two-qubit gates equal in number up to additional single-qubit rotations, the basis set has less impact on final results than the choice of topology. SDK-specific settings, such as optimization level, can also be set in this file, and additional options can be easily added as there is no hard coding of parameters. In addition, \texttt{pytest} and \texttt{pytest-benchmark} options can be set using the standard \texttt{pytest.ini} file. In this file, we add the flag to allow for only a single execution of a test to be performed, as opposed to the usual minimum of five to make runtimes manageable.

\subsection*{SDK specific considerations}
Given quantum computing software's nascent stage of development, it is common to encounter pitfalls when benchmarking these software stacks. Here, we detail some of these issues as they pertain to executing tests in Benchpress.

\subsubsection*{BQSKit}
BQSKit performs unitary synthesis up to a maximum size specified by the \texttt{max\_synthesis\_size} argument to the compiler. The compiler will fail if a unitary is larger than this value. However, given an OpenQASM circuit, an end user must first parse the file to learn the correct size for this argument; the returned error message does not include the required value. As there is no manner outside of parsing files to gain this information, we have this parameter to the default value, \texttt{max\_synthesis\_size=3}, letting tests fail if they have unitary gates outside of this value.

In addition, BQSKit does not support coupling maps that correspond to two-qubit entangling gates with directionality; the topology is assumed to be symmetric.  The $CZ$ gate used in this work is a symmetric gate, and thus the circuits returned from the BQSKit transpiler pass the structural validation performed here.  Selecting a directional gate, such as an echoed cross-resonance (ECR) gate, would in general fail validation.

\subsubsection*{Qiskit Transpiler Service}

The default timeout value for the QTS is less than the $3600$ second timeout used in this work.  As such, we explicitly set the timeout value to match when calling the QTS service for each test.

\subsubsection*{Tket}
The OpenQASM import functionality in Tket requires the user to specify the size of classical registers in the circuit if those registers are larger than $32$ bits.  Given an arbitrary OpenQASM file, the user must first parse the file to gather the size of the classical registers or try importing the file first, capturing the exception, and reading the register size from the error message.  To get around this limitation, Benchpress includes a \texttt{maxwidth} parameter in the Tket section of the \texttt{default.conf} file that allows for specifying a maximum allowed classical register size.  Given that the maximum number of qubits in the OpenQASM files is $433$, we have set this value to $500$ in the \texttt{default.conf}.

\subsubsection*{Staq}\label{sec:staq}
Staq cannot return quantum circuits in the basis set of the target backend.  Instead, the output is always expressed in generic one-qubit unitary $U$ and CNOT gates.  Because of this, we only perform structural validation on circuits returned by Staq.  In addition, we compute two-qubit gate counts and depth on the CNOT gates.  This is valid provided the target two-qubit gate is equivalent to a CNOT gate up to single-qubit rotations.  For the CZ gate used here, this relation holds.

Optimization level $3$ of Staq includes the compiler flag \texttt{-c} that applies a CNOT optimization pass.  This pass generates output OpenQASM files that do not obey the entangling gate topology of the target device; the output circuits fail the structural validation check at the end of each test.  As such, we have set the default optimization level of Staq to $2$.

\section{Data availability}\label{sec:data}

Results generated in this study are included in the \texttt{published\_results} directory at: \href{https://github.com/Qiskit/benchpress/tree/1.0}{https://github.com/Qiskit/benchpress/tree/1.0}.

\section{Code availability}\label{sec:code}

All code used in generating this data set is open-source and available at Ref.~(\onlinecite{benchpress}).

\section*{Author contributions}
P.D.N devised the testing infrastructure and methodology, as well as contributed to the code base.  A.A.S. wrote the tests for both the BQSKit and Staq SDKs and contributed to the infrastructure code.  S.B. wrote the code for Hamiltonian simulation tests from HamLib, and selected the subset of Hamiltonians that are included in this work.  L.B. added the test logic that allows for skipping tests with timeout logic, while S.G added random Clifford tests.  M.T. added the Feynman suite of tests for the device transpilation tests, and A.J.A suggested circuit libraries for inclusion and test cases to consider.  All authors contributed to the manuscript.

\section*{Competing interests}
The authors declare no competing interests.

\begin{acknowledgments}

The authors thank Julien Gacon, Alexander Ivrii, and Jake Lishman for helpful discussions.
L.B., M.T., and A.J-A were supported by the U.S. Department of Energy, Office of Science, National Quantum Information Science Research Centers, Co-design Center for Quantum Advantage (C2QA) under contract number DE-SC0012704.

\end{acknowledgments}

\bibliography{refs}

\begin{thebibliography}{51}%
\makeatletter
\providecommand \@ifxundefined [1]{%
 \@ifx{#1\undefined}
}%
\providecommand \@ifnum [1]{%
 \ifnum #1\expandafter \@firstoftwo
 \else \expandafter \@secondoftwo
 \fi
}%
\providecommand \@ifx [1]{%
 \ifx #1\expandafter \@firstoftwo
 \else \expandafter \@secondoftwo
 \fi
}%
\providecommand \natexlab [1]{#1}%
\providecommand \enquote  [1]{``#1''}%
\providecommand \bibnamefont  [1]{#1}%
\providecommand \bibfnamefont [1]{#1}%
\providecommand \citenamefont [1]{#1}%
\providecommand \href@noop [0]{\@secondoftwo}%
\providecommand \href [0]{\begingroup \@sanitize@url \@href}%
\providecommand \@href[1]{\@@startlink{#1}\@@href}%
\providecommand \@@href[1]{\endgroup#1\@@endlink}%
\providecommand \@sanitize@url [0]{\catcode `\\12\catcode `\$12\catcode
  `\&12\catcode `\#12\catcode `\^12\catcode `\_12\catcode `\%12\relax}%
\providecommand \@@startlink[1]{}%
\providecommand \@@endlink[0]{}%
\providecommand \url  [0]{\begingroup\@sanitize@url \@url }%
\providecommand \@url [1]{\endgroup\@href {#1}{\urlprefix }}%
\providecommand \urlprefix  [0]{URL }%
\providecommand \Eprint [0]{\href }%
\providecommand \doibase [0]{https://doi.org/}%
\providecommand \selectlanguage [0]{\@gobble}%
\providecommand \bibinfo  [0]{\@secondoftwo}%
\providecommand \bibfield  [0]{\@secondoftwo}%
\providecommand \translation [1]{[#1]}%
\providecommand \BibitemOpen [0]{}%
\providecommand \bibitemStop [0]{}%
\providecommand \bibitemNoStop [0]{.\EOS\space}%
\providecommand \EOS [0]{\spacefactor3000\relax}%
\providecommand \BibitemShut  [1]{\csname bibitem#1\endcsname}%
\let\auto@bib@innerbib\@empty
\bibitem [{\citenamefont {Services}(2020)}]{braket:2020}%
  \BibitemOpen
  \bibfield  {author} {\bibinfo {author} {\bibfnamefont {A.~W.}\ \bibnamefont
  {Services}},\ }\href {https://aws.amazon.com/braket/} {\bibinfo {title}
  {{A}mazon {B}raket}} (\bibinfo {year} {2020})\BibitemShut {NoStop}%
\bibitem [{\citenamefont {Younis}\ \emph {et~al.}(2021)\citenamefont {Younis},
  \citenamefont {Iancu}, \citenamefont {Lavrijsen}, \citenamefont {Davis},
  \citenamefont {Smith},\ and\ \citenamefont {USDOE}}]{younis:2021}%
  \BibitemOpen
  \bibfield  {author} {\bibinfo {author} {\bibfnamefont {E.}~\bibnamefont
  {Younis}}, \bibinfo {author} {\bibfnamefont {C.~C.}\ \bibnamefont {Iancu}},
  \bibinfo {author} {\bibfnamefont {W.}~\bibnamefont {Lavrijsen}}, \bibinfo
  {author} {\bibfnamefont {M.}~\bibnamefont {Davis}}, \bibinfo {author}
  {\bibfnamefont {E.}~\bibnamefont {Smith}},\ and\ \bibinfo {author}
  {\bibnamefont {USDOE}},\ }\href {https://doi.org/10.11578/dc.20210603.2}
  {\bibinfo {title} {Berkeley quantum synthesis toolkit (bqskit) v1}} (\bibinfo
  {year} {2021})\BibitemShut {NoStop}%
\bibitem [{\citenamefont {Developers}()}]{cirq}%
  \BibitemOpen
  \bibfield  {author} {\bibinfo {author} {\bibfnamefont {C.}~\bibnamefont
  {Developers}},\ }\href {https://doi.org/10.5281/zenodo.4062499} {\bibinfo
  {title} {{C}irq}}\BibitemShut {NoStop}%
\bibitem [{\citenamefont {Javadi-Abhari}\ \emph {et~al.}(2024)\citenamefont
  {Javadi-Abhari}, \citenamefont {Treinish}, \citenamefont {Krsulich},
  \citenamefont {Wood}, \citenamefont {Lishman}, \citenamefont {Gacon},
  \citenamefont {Martiel}, \citenamefont {Nation}, \citenamefont {Bishop},
  \citenamefont {Cross}, \citenamefont {Johnson},\ and\ \citenamefont
  {Gambetta}}]{javadi:2024}%
  \BibitemOpen
  \bibfield  {author} {\bibinfo {author} {\bibfnamefont {A.}~\bibnamefont
  {Javadi-Abhari}}, \bibinfo {author} {\bibfnamefont {M.}~\bibnamefont
  {Treinish}}, \bibinfo {author} {\bibfnamefont {K.}~\bibnamefont {Krsulich}},
  \bibinfo {author} {\bibfnamefont {C.~J.}\ \bibnamefont {Wood}}, \bibinfo
  {author} {\bibfnamefont {J.}~\bibnamefont {Lishman}}, \bibinfo {author}
  {\bibfnamefont {J.}~\bibnamefont {Gacon}}, \bibinfo {author} {\bibfnamefont
  {S.}~\bibnamefont {Martiel}}, \bibinfo {author} {\bibfnamefont {P.~D.}\
  \bibnamefont {Nation}}, \bibinfo {author} {\bibfnamefont {L.~S.}\
  \bibnamefont {Bishop}}, \bibinfo {author} {\bibfnamefont {A.~W.}\
  \bibnamefont {Cross}}, \bibinfo {author} {\bibfnamefont {B.~R.}\ \bibnamefont
  {Johnson}},\ and\ \bibinfo {author} {\bibfnamefont {J.~M.}\ \bibnamefont
  {Gambetta}},\ }\bibfield  {title} {\bibinfo {title} {{Q}uantum computing with
  {Q}iskit},\ }\bibfield  {journal} {\bibinfo  {journal} {arXiv:2405.08810}\
  }\href {https://doi.org/10.48550/arXiv.2405.08810}
  {10.48550/arXiv.2405.08810} (\bibinfo {year} {2024})\BibitemShut {NoStop}%
\bibitem [{\citenamefont {Kremer}\ \emph {et~al.}(2024)\citenamefont {Kremer},
  \citenamefont {Villar}, \citenamefont {Paik}, \citenamefont {Duran},
  \citenamefont {Faro},\ and\ \citenamefont {Crus-Benito}}]{kremer:2024}%
  \BibitemOpen
  \bibfield  {author} {\bibinfo {author} {\bibfnamefont {D.}~\bibnamefont
  {Kremer}}, \bibinfo {author} {\bibfnamefont {V.}~\bibnamefont {Villar}},
  \bibinfo {author} {\bibfnamefont {H.}~\bibnamefont {Paik}}, \bibinfo {author}
  {\bibfnamefont {I.}~\bibnamefont {Duran}}, \bibinfo {author} {\bibfnamefont
  {I.}~\bibnamefont {Faro}},\ and\ \bibinfo {author} {\bibfnamefont
  {J.}~\bibnamefont {Crus-Benito}},\ }\bibfield  {title} {\bibinfo {title}
  {{P}ractical and efficient quantum circuit synthesis and transpiling with
  {R}einforcement {L}earning},\ }\bibfield  {journal} {\bibinfo  {journal}
  {arXiv:2405.13196}\ }\href {https://doi.org/10.48550/arXiv.2405.13196}
  {10.48550/arXiv.2405.13196} (\bibinfo {year} {2024})\BibitemShut {NoStop}%
\bibitem [{\citenamefont {Amy}\ and\ \citenamefont
  {Gheorghiu}(2020)}]{amy:2020}%
  \BibitemOpen
  \bibfield  {author} {\bibinfo {author} {\bibfnamefont {M.}~\bibnamefont
  {Amy}}\ and\ \bibinfo {author} {\bibfnamefont {V.}~\bibnamefont
  {Gheorghiu}},\ }\bibfield  {title} {\bibinfo {title} {staq---a full-stack
  quantum processing toolkit},\ }\href
  {https://doi.org/10.1088/2058-9565/ab9359} {\bibfield  {journal} {\bibinfo
  {journal} {Quantum Science and Technology}\ }\textbf {\bibinfo {volume}
  {5}},\ \bibinfo {pages} {034016} (\bibinfo {year} {2020})}\BibitemShut
  {NoStop}%
\bibitem [{\citenamefont {Sivarajah}\ \emph {et~al.}(2020)\citenamefont
  {Sivarajah}, \citenamefont {Dilkes}, \citenamefont {Cowtan}, \citenamefont
  {Simmons}, \citenamefont {Edgington},\ and\ \citenamefont
  {Duncan}}]{sivarajah:2021}%
  \BibitemOpen
  \bibfield  {author} {\bibinfo {author} {\bibfnamefont {S.}~\bibnamefont
  {Sivarajah}}, \bibinfo {author} {\bibfnamefont {S.}~\bibnamefont {Dilkes}},
  \bibinfo {author} {\bibfnamefont {A.}~\bibnamefont {Cowtan}}, \bibinfo
  {author} {\bibfnamefont {W.}~\bibnamefont {Simmons}}, \bibinfo {author}
  {\bibfnamefont {A.}~\bibnamefont {Edgington}},\ and\ \bibinfo {author}
  {\bibfnamefont {R.}~\bibnamefont {Duncan}},\ }\bibfield  {title} {\bibinfo
  {title} {t|ket⟩: a retargetable compiler for nisq devices},\ }\href
  {https://doi.org/10.1088/2058-9565/ab8e92} {\bibfield  {journal} {\bibinfo
  {journal} {Quantum Science and Technology}\ }\textbf {\bibinfo {volume}
  {6}},\ \bibinfo {pages} {014003} (\bibinfo {year} {2020})}\BibitemShut
  {NoStop}%
\bibitem [{\citenamefont {Amy}(2019)}]{amy:2019}%
  \BibitemOpen
  \bibfield  {author} {\bibinfo {author} {\bibfnamefont {M.}~\bibnamefont
  {Amy}},\ }\bibfield  {title} {\bibinfo {title} {Towards large-scale
  functional verification of universal quantum circuits},\ }in\ \href
  {https://doi.org/10.4204/EPTCS.287.1} {\emph {\bibinfo {booktitle}
  {Proceedings of the 15th International Conference on Quantum Physics and
  Logic}}}\ (\bibinfo {year} {2019})\ pp.\ \bibinfo {pages} {1--21}\BibitemShut
  {NoStop}%
\bibitem [{\citenamefont {Li}\ \emph {et~al.}(2023)\citenamefont {Li},
  \citenamefont {Stein}, \citenamefont {Krishnamoorthy},\ and\ \citenamefont
  {Ang}}]{li:2023}%
  \BibitemOpen
  \bibfield  {author} {\bibinfo {author} {\bibfnamefont {A.}~\bibnamefont
  {Li}}, \bibinfo {author} {\bibfnamefont {S.}~\bibnamefont {Stein}}, \bibinfo
  {author} {\bibfnamefont {S.}~\bibnamefont {Krishnamoorthy}},\ and\ \bibinfo
  {author} {\bibfnamefont {J.}~\bibnamefont {Ang}},\ }\bibfield  {title}
  {\bibinfo {title} {Qasmbench: A low-level quantum benchmark suite for nisq
  evaluation and simulation},\ }\bibfield  {journal} {\bibinfo  {journal} {ACM
  Transactions on Quantum Computing}\ }\textbf {\bibinfo {volume} {4}},\ \href
  {https://doi.org/10.1145/3550488} {10.1145/3550488} (\bibinfo {year}
  {2023})\BibitemShut {NoStop}%
\bibitem [{\citenamefont {Mori}\ \emph {et~al.}(2023)\citenamefont {Mori},
  \citenamefont {Hakoshima}, \citenamefont {Sudo}, \citenamefont {Mori},
  \citenamefont {Mitarai},\ and\ \citenamefont {Fujii}}]{mori:2023}%
  \BibitemOpen
  \bibfield  {author} {\bibinfo {author} {\bibfnamefont {Y.}~\bibnamefont
  {Mori}}, \bibinfo {author} {\bibfnamefont {H.}~\bibnamefont {Hakoshima}},
  \bibinfo {author} {\bibfnamefont {K.}~\bibnamefont {Sudo}}, \bibinfo {author}
  {\bibfnamefont {T.}~\bibnamefont {Mori}}, \bibinfo {author} {\bibfnamefont
  {K.}~\bibnamefont {Mitarai}},\ and\ \bibinfo {author} {\bibfnamefont
  {K.}~\bibnamefont {Fujii}},\ }\bibfield  {title} {\bibinfo {title} {Quantum
  circuit unoptimization},\ }\bibfield  {journal} {\bibinfo  {journal}
  {arXiv:2311.03805}\ }\href {https://doi.org/10.48550/arXiv.2311.03805}
  {10.48550/arXiv.2311.03805} (\bibinfo {year} {2023})\BibitemShut {NoStop}%
\bibitem [{\citenamefont {Sawaya}\ \emph {et~al.}(2023)\citenamefont {Sawaya},
  \citenamefont {Marti-Dafcik}, \citenamefont {Ho}, \citenamefont {Tabor},
  \citenamefont {Bernal~Neira}, \citenamefont {Magann}, \citenamefont
  {Premaratne}, \citenamefont {Dubey}, \citenamefont {Matsuura}, \citenamefont
  {Bishop}, \citenamefont {De~Jong}, \citenamefont {Benjamin}, \citenamefont
  {Parekh}, \citenamefont {Tubman}, \citenamefont {Klymko},\ and\ \citenamefont
  {Camps}}]{sawaya:2023}%
  \BibitemOpen
  \bibfield  {author} {\bibinfo {author} {\bibfnamefont {N.~P.}\ \bibnamefont
  {Sawaya}}, \bibinfo {author} {\bibfnamefont {D.}~\bibnamefont
  {Marti-Dafcik}}, \bibinfo {author} {\bibfnamefont {Y.}~\bibnamefont {Ho}},
  \bibinfo {author} {\bibfnamefont {D.~P.}\ \bibnamefont {Tabor}}, \bibinfo
  {author} {\bibfnamefont {D.~E.}\ \bibnamefont {Bernal~Neira}}, \bibinfo
  {author} {\bibfnamefont {A.~B.}\ \bibnamefont {Magann}}, \bibinfo {author}
  {\bibfnamefont {S.}~\bibnamefont {Premaratne}}, \bibinfo {author}
  {\bibfnamefont {P.}~\bibnamefont {Dubey}}, \bibinfo {author} {\bibfnamefont
  {A.}~\bibnamefont {Matsuura}}, \bibinfo {author} {\bibfnamefont
  {N.}~\bibnamefont {Bishop}}, \bibinfo {author} {\bibfnamefont {W.~A.}\
  \bibnamefont {De~Jong}}, \bibinfo {author} {\bibfnamefont {S.}~\bibnamefont
  {Benjamin}}, \bibinfo {author} {\bibfnamefont {O.~D.}\ \bibnamefont
  {Parekh}}, \bibinfo {author} {\bibfnamefont {N.~M.}\ \bibnamefont {Tubman}},
  \bibinfo {author} {\bibfnamefont {K.}~\bibnamefont {Klymko}},\ and\ \bibinfo
  {author} {\bibfnamefont {D.}~\bibnamefont {Camps}},\ }\bibfield  {title}
  {\bibinfo {title} {{H}am{L}ib: {A} {L}ibrary of {H}amiltonians for
  {B}enchmarking {Q}uantum {A}lgorithms and {H}ardware},\ }in\ \href
  {https://doi.org/10.1109/QCE57702.2023.10296} {\emph {\bibinfo {booktitle}
  {2023 IEEE International Conference on Quantum Computing and Engineering
  (QCE)}}},\ Vol.~\bibinfo {volume} {2}\ (\bibinfo {year} {2023})\ p.\ \bibinfo
  {pages} {389}\BibitemShut {NoStop}%
\bibitem [{\citenamefont {Kharkov}\ \emph {et~al.}(2022)\citenamefont
  {Kharkov}, \citenamefont {Ivanova}, \citenamefont {Mikhantiev},\ and\
  \citenamefont {Kotelnikov}}]{kharkov:2022}%
  \BibitemOpen
  \bibfield  {author} {\bibinfo {author} {\bibfnamefont {Y.}~\bibnamefont
  {Kharkov}}, \bibinfo {author} {\bibfnamefont {A.}~\bibnamefont {Ivanova}},
  \bibinfo {author} {\bibfnamefont {E.}~\bibnamefont {Mikhantiev}},\ and\
  \bibinfo {author} {\bibfnamefont {A.}~\bibnamefont {Kotelnikov}},\ }\bibfield
   {title} {\bibinfo {title} {{A}rline {B}enchmarks: {A}utomated {B}enchmarking
  {P}latform for {Q}uantum {C}ompilers},\ }\bibfield  {journal} {\bibinfo
  {journal} {arXiv:2202.14025}\ }\href
  {https://doi.org/10.48550/arXiv.2202.14025} {10.48550/arXiv.2202.14025}
  (\bibinfo {year} {2022})\BibitemShut {NoStop}%
\bibitem [{\citenamefont {Quetschlich}\ \emph {et~al.}(2023)\citenamefont
  {Quetschlich}, \citenamefont {Burgholzer},\ and\ \citenamefont
  {Wille}}]{quetschlich:2023}%
  \BibitemOpen
  \bibfield  {author} {\bibinfo {author} {\bibfnamefont {N.}~\bibnamefont
  {Quetschlich}}, \bibinfo {author} {\bibfnamefont {L.}~\bibnamefont
  {Burgholzer}},\ and\ \bibinfo {author} {\bibfnamefont {R.}~\bibnamefont
  {Wille}},\ }\bibfield  {title} {\bibinfo {title} {{MQT} {B}ench:
  {B}enchmarking {S}oftware and {D}esign {A}utomation {T}ools for {Q}uantum
  {C}omputing},\ }\href {https://doi.org/10.22331/q-2023-07-20-1062} {\bibfield
   {journal} {\bibinfo  {journal} {{Quantum}}\ }\textbf {\bibinfo {volume}
  {7}},\ \bibinfo {pages} {2062} (\bibinfo {year} {2023})}\BibitemShut
  {NoStop}%
\bibitem [{\citenamefont {Cross}\ \emph {et~al.}(2022)\citenamefont {Cross},
  \citenamefont {Javadi-Abhari}, \citenamefont {Alexander}, \citenamefont
  {De~Beaudrap}, \citenamefont {Bishop}, \citenamefont {Heidel}, \citenamefont
  {Ryan}, \citenamefont {Sivarajah}, \citenamefont {Smolin}, \citenamefont
  {Gambetta} \emph {et~al.}}]{cross:2022}%
  \BibitemOpen
  \bibfield  {author} {\bibinfo {author} {\bibfnamefont {A.}~\bibnamefont
  {Cross}}, \bibinfo {author} {\bibfnamefont {A.}~\bibnamefont
  {Javadi-Abhari}}, \bibinfo {author} {\bibfnamefont {T.}~\bibnamefont
  {Alexander}}, \bibinfo {author} {\bibfnamefont {N.}~\bibnamefont
  {De~Beaudrap}}, \bibinfo {author} {\bibfnamefont {L.~S.}\ \bibnamefont
  {Bishop}}, \bibinfo {author} {\bibfnamefont {S.}~\bibnamefont {Heidel}},
  \bibinfo {author} {\bibfnamefont {C.~A.}\ \bibnamefont {Ryan}}, \bibinfo
  {author} {\bibfnamefont {P.}~\bibnamefont {Sivarajah}}, \bibinfo {author}
  {\bibfnamefont {J.}~\bibnamefont {Smolin}}, \bibinfo {author} {\bibfnamefont
  {J.~M.}\ \bibnamefont {Gambetta}}, \emph {et~al.},\ }\bibfield  {title}
  {\bibinfo {title} {Openqasm 3: A broader and deeper quantum assembly
  language},\ }\href@noop {} {\bibfield  {journal} {\bibinfo  {journal} {ACM
  Transactions on Quantum Computing}\ }\textbf {\bibinfo {volume} {3}},\
  \bibinfo {pages} {1} (\bibinfo {year} {2022})}\BibitemShut {NoStop}%
\bibitem [{\citenamefont {Yan}\ \emph {et~al.}(2024)\citenamefont {Yan},
  \citenamefont {Wu}, \citenamefont {Chen}, \citenamefont {Pan}, \citenamefont
  {Lu}, \citenamefont {Zhou}, \citenamefont {Wang}, \citenamefont {Wang},\ and\
  \citenamefont {Yan}}]{yan:2024}%
  \BibitemOpen
  \bibfield  {author} {\bibinfo {author} {\bibfnamefont {G.}~\bibnamefont
  {Yan}}, \bibinfo {author} {\bibfnamefont {W.}~\bibnamefont {Wu}}, \bibinfo
  {author} {\bibfnamefont {Y.}~\bibnamefont {Chen}}, \bibinfo {author}
  {\bibfnamefont {K.}~\bibnamefont {Pan}}, \bibinfo {author} {\bibfnamefont
  {X.}~\bibnamefont {Lu}}, \bibinfo {author} {\bibfnamefont {Z.}~\bibnamefont
  {Zhou}}, \bibinfo {author} {\bibfnamefont {Y.}~\bibnamefont {Wang}}, \bibinfo
  {author} {\bibfnamefont {R.}~\bibnamefont {Wang}},\ and\ \bibinfo {author}
  {\bibfnamefont {J.}~\bibnamefont {Yan}},\ }\bibfield  {title} {\bibinfo
  {title} {{Q}uantum {C}ircuit {S}ynthesis and {C}ompilation {O}ptimization:
  {O}verview and {P}rospects},\ }\bibfield  {journal} {\bibinfo  {journal}
  {arXiv:2407.00736}\ }\href {https://doi.org/10.48550/arXiv.2407.00736}
  {10.48550/arXiv.2407.00736} (\bibinfo {year} {2024})\BibitemShut {NoStop}%
\bibitem [{\citenamefont {Zhang}\ and\ \citenamefont
  {Nation}(2023)}]{zhang:2023}%
  \BibitemOpen
  \bibfield  {author} {\bibinfo {author} {\bibfnamefont {V.}~\bibnamefont
  {Zhang}}\ and\ \bibinfo {author} {\bibfnamefont {P.~D.}\ \bibnamefont
  {Nation}},\ }\bibfield  {title} {\bibinfo {title} {{C}haracterizing quantum
  processors using discrete time crystals},\ }\bibfield  {journal} {\bibinfo
  {journal} {arXiv:2301.07625}\ }\href
  {https://doi.org/10.48550/arXiv.2301.07625} {10.48550/arXiv.2301.07625}
  (\bibinfo {year} {2023})\BibitemShut {NoStop}%
\bibitem [{\citenamefont {Kim}\ \emph {et~al.}(2023)\citenamefont {Kim},
  \citenamefont {Eddins}, \citenamefont {Anand}, \citenamefont {Wei},
  \citenamefont {van~den Berg}, \citenamefont {Rosenblatt}, \citenamefont
  {Nayfeh}, \citenamefont {Wu}, \citenamefont {Zaletel}, \citenamefont
  {Temme},\ and\ \citenamefont {Kandala}}]{kim:2023}%
  \BibitemOpen
  \bibfield  {author} {\bibinfo {author} {\bibfnamefont {Y.}~\bibnamefont
  {Kim}}, \bibinfo {author} {\bibfnamefont {A.}~\bibnamefont {Eddins}},
  \bibinfo {author} {\bibfnamefont {S.}~\bibnamefont {Anand}}, \bibinfo
  {author} {\bibfnamefont {K.~X.}\ \bibnamefont {Wei}}, \bibinfo {author}
  {\bibfnamefont {E.}~\bibnamefont {van~den Berg}}, \bibinfo {author}
  {\bibfnamefont {S.}~\bibnamefont {Rosenblatt}}, \bibinfo {author}
  {\bibfnamefont {H.}~\bibnamefont {Nayfeh}}, \bibinfo {author} {\bibfnamefont
  {Y.}~\bibnamefont {Wu}}, \bibinfo {author} {\bibfnamefont {M.}~\bibnamefont
  {Zaletel}}, \bibinfo {author} {\bibfnamefont {K.}~\bibnamefont {Temme}},\
  and\ \bibinfo {author} {\bibfnamefont {A.}~\bibnamefont {Kandala}},\
  }\bibfield  {title} {\bibinfo {title} {Evidence for the utility of quantum
  computing before fault tolerance},\ }\href
  {https://doi.org/10.1038/s41586-023-06096-3} {\bibfield  {journal} {\bibinfo
  {journal} {Nature}\ }\textbf {\bibinfo {volume} {618}},\ \bibinfo {pages}
  {500} (\bibinfo {year} {2023})}\BibitemShut {NoStop}%
\bibitem [{\citenamefont {Yu}\ \emph {et~al.}(2023)\citenamefont {Yu},
  \citenamefont {Zhao},\ and\ \citenamefont {Wei}}]{hongye:2023}%
  \BibitemOpen
  \bibfield  {author} {\bibinfo {author} {\bibfnamefont {H.}~\bibnamefont
  {Yu}}, \bibinfo {author} {\bibfnamefont {Y.}~\bibnamefont {Zhao}},\ and\
  \bibinfo {author} {\bibfnamefont {T.-C.}\ \bibnamefont {Wei}},\ }\bibfield
  {title} {\bibinfo {title} {Simulating large-size quantum spin chains on
  cloud-based superconducting quantum computers},\ }\href
  {https://doi.org/10.1103/PhysRevResearch.5.013183} {\bibfield  {journal}
  {\bibinfo  {journal} {Phys. Rev. Res.}\ }\textbf {\bibinfo {volume} {5}},\
  \bibinfo {pages} {013183} (\bibinfo {year} {2023})}\BibitemShut {NoStop}%
\bibitem [{\citenamefont {Majumdar}\ \emph {et~al.}(2023)\citenamefont
  {Majumdar}, \citenamefont {Rivero}, \citenamefont {Metz}, \citenamefont
  {Hasan},\ and\ \citenamefont {Wang}}]{majumdar:2023}%
  \BibitemOpen
  \bibfield  {author} {\bibinfo {author} {\bibfnamefont {R.}~\bibnamefont
  {Majumdar}}, \bibinfo {author} {\bibfnamefont {P.}~\bibnamefont {Rivero}},
  \bibinfo {author} {\bibfnamefont {F.}~\bibnamefont {Metz}}, \bibinfo {author}
  {\bibfnamefont {A.}~\bibnamefont {Hasan}},\ and\ \bibinfo {author}
  {\bibfnamefont {D.~S.}\ \bibnamefont {Wang}},\ }\bibfield  {title} {\bibinfo
  {title} {{B}est practices for quantum error mitigation with digital
  zero-noise extrapolation},\ }\bibfield  {journal} {\bibinfo  {journal}
  {arXiv:2307.05203}\ }\href {https://doi.org/10.48550/arXiv.2307.05203}
  {10.48550/arXiv.2307.05203} (\bibinfo {year} {2023})\BibitemShut {NoStop}%
\bibitem [{\citenamefont {Shtanko}\ \emph {et~al.}(2023)\citenamefont
  {Shtanko}, \citenamefont {Wang}, \citenamefont {Zhang}, \citenamefont
  {Harle}, \citenamefont {Seif}, \citenamefont {Movassagh},\ and\ \citenamefont
  {Minev}}]{shtanko:2023}%
  \BibitemOpen
  \bibfield  {author} {\bibinfo {author} {\bibfnamefont {O.}~\bibnamefont
  {Shtanko}}, \bibinfo {author} {\bibfnamefont {D.~S.}\ \bibnamefont {Wang}},
  \bibinfo {author} {\bibfnamefont {H.}~\bibnamefont {Zhang}}, \bibinfo
  {author} {\bibfnamefont {N.}~\bibnamefont {Harle}}, \bibinfo {author}
  {\bibfnamefont {A.}~\bibnamefont {Seif}}, \bibinfo {author} {\bibfnamefont
  {R.}~\bibnamefont {Movassagh}},\ and\ \bibinfo {author} {\bibfnamefont
  {Z.}~\bibnamefont {Minev}},\ }\bibfield  {title} {\bibinfo {title}
  {{U}ncovering {L}ocal {I}ntegrability in {Q}uantum {M}any-{B}ody
  {D}ynamics},\ }\bibfield  {journal} {\bibinfo  {journal} {arXiv:2307.07552}\
  }\href {https://doi.org/10.48550/arXiv.2307.07552}
  {10.48550/arXiv.2307.07552} (\bibinfo {year} {2023})\BibitemShut {NoStop}%
\bibitem [{\citenamefont {Yasuda}\ \emph {et~al.}(2023)\citenamefont {Yasuda},
  \citenamefont {Suzuki}, \citenamefont {Kubota}, \citenamefont {Nakajima},
  \citenamefont {Gao}, \citenamefont {Zhang}, \citenamefont {Shimono},
  \citenamefont {Nurdin},\ and\ \citenamefont {Yamamoto}}]{yasuda:2023}%
  \BibitemOpen
  \bibfield  {author} {\bibinfo {author} {\bibfnamefont {T.}~\bibnamefont
  {Yasuda}}, \bibinfo {author} {\bibfnamefont {Y.}~\bibnamefont {Suzuki}},
  \bibinfo {author} {\bibfnamefont {T.}~\bibnamefont {Kubota}}, \bibinfo
  {author} {\bibfnamefont {K.}~\bibnamefont {Nakajima}}, \bibinfo {author}
  {\bibfnamefont {Q.}~\bibnamefont {Gao}}, \bibinfo {author} {\bibfnamefont
  {W.}~\bibnamefont {Zhang}}, \bibinfo {author} {\bibfnamefont
  {S.}~\bibnamefont {Shimono}}, \bibinfo {author} {\bibfnamefont {H.~I.}\
  \bibnamefont {Nurdin}},\ and\ \bibinfo {author} {\bibfnamefont
  {N.}~\bibnamefont {Yamamoto}},\ }\bibfield  {title} {\bibinfo {title}
  {{Q}uantum reservoir computing with repeated measurements on superconducting
  devices},\ }\bibfield  {journal} {\bibinfo  {journal} {arXiv:2310.06706}\
  }\href {https://doi.org/10.48550/arXiv.2310.06706}
  {10.48550/arXiv.2310.06706} (\bibinfo {year} {2023})\BibitemShut {NoStop}%
\bibitem [{\citenamefont {Chen}\ \emph {et~al.}(2023)\citenamefont {Chen},
  \citenamefont {Zhu}, \citenamefont {Verresen}, \citenamefont {Seif},
  \citenamefont {B\"{a}umer}, \citenamefont {Layden}, \citenamefont
  {Tantivasadakarn}, \citenamefont {Zhu}, \citenamefont {Sheldon},
  \citenamefont {Vishwanath}, \citenamefont {Trebst},\ and\ \citenamefont
  {Kandala}}]{chen:2023}%
  \BibitemOpen
  \bibfield  {author} {\bibinfo {author} {\bibfnamefont {E.~H.}\ \bibnamefont
  {Chen}}, \bibinfo {author} {\bibfnamefont {G.-Y.}\ \bibnamefont {Zhu}},
  \bibinfo {author} {\bibfnamefont {R.}~\bibnamefont {Verresen}}, \bibinfo
  {author} {\bibfnamefont {A.}~\bibnamefont {Seif}}, \bibinfo {author}
  {\bibfnamefont {E.}~\bibnamefont {B\"{a}umer}}, \bibinfo {author}
  {\bibfnamefont {D.}~\bibnamefont {Layden}}, \bibinfo {author} {\bibfnamefont
  {N.}~\bibnamefont {Tantivasadakarn}}, \bibinfo {author} {\bibfnamefont
  {G.}~\bibnamefont {Zhu}}, \bibinfo {author} {\bibfnamefont {S.}~\bibnamefont
  {Sheldon}}, \bibinfo {author} {\bibfnamefont {A.}~\bibnamefont {Vishwanath}},
  \bibinfo {author} {\bibfnamefont {S.}~\bibnamefont {Trebst}},\ and\ \bibinfo
  {author} {\bibfnamefont {A.}~\bibnamefont {Kandala}},\ }\bibfield  {title}
  {\bibinfo {title} {{R}ealizing the {N}ishimori transition across the error
  threshold for constant-depth quantum circuits},\ }\bibfield  {journal}
  {\bibinfo  {journal} {arXiv:2309.02863}\ }\href
  {https://doi.org/10.48550/arXiv.2309.02863} {10.48550/arXiv.2309.02863}
  (\bibinfo {year} {2023})\BibitemShut {NoStop}%
\bibitem [{\citenamefont {Farrell}\ \emph
  {et~al.}(2024{\natexlab{a}})\citenamefont {Farrell}, \citenamefont {Illa},
  \citenamefont {Ciavarella},\ and\ \citenamefont {Savage}}]{farrell:2024}%
  \BibitemOpen
  \bibfield  {author} {\bibinfo {author} {\bibfnamefont {R.~C.}\ \bibnamefont
  {Farrell}}, \bibinfo {author} {\bibfnamefont {M.}~\bibnamefont {Illa}},
  \bibinfo {author} {\bibfnamefont {A.~N.}\ \bibnamefont {Ciavarella}},\ and\
  \bibinfo {author} {\bibfnamefont {M.~J.}\ \bibnamefont {Savage}},\ }\bibfield
   {title} {\bibinfo {title} {Scalable circuits for preparing ground states on
  digital quantum computers: The schwinger model vacuum on 100 qubits},\ }\href
  {https://doi.org/10.1103/PRXQuantum.5.020315} {\bibfield  {journal} {\bibinfo
   {journal} {PRX Quantum}\ }\textbf {\bibinfo {volume} {5}},\ \bibinfo {pages}
  {020315} (\bibinfo {year} {2024}{\natexlab{a}})}\BibitemShut {NoStop}%
\bibitem [{\citenamefont {Pelofske}\ \emph {et~al.}(2023)\citenamefont
  {Pelofske}, \citenamefont {B\"{a}rtschi}, \citenamefont {Cincio},
  \citenamefont {Golden},\ and\ \citenamefont {Eidenbenz}}]{pelofske:2023}%
  \BibitemOpen
  \bibfield  {author} {\bibinfo {author} {\bibfnamefont {E.}~\bibnamefont
  {Pelofske}}, \bibinfo {author} {\bibfnamefont {A.}~\bibnamefont
  {B\"{a}rtschi}}, \bibinfo {author} {\bibfnamefont {L.}~\bibnamefont
  {Cincio}}, \bibinfo {author} {\bibfnamefont {J.}~\bibnamefont {Golden}},\
  and\ \bibinfo {author} {\bibfnamefont {S.}~\bibnamefont {Eidenbenz}},\
  }\bibfield  {title} {\bibinfo {title} {{S}caling {W}hole-{C}hip {QAOA} for
  {H}igher-{O}rder {I}sing {S}pin {G}lass {M}odels on {H}eavy-{H}ex {G}raphs},\
  }\bibfield  {journal} {\bibinfo  {journal} {arXiv:2312.00997}\ }\href
  {https://doi.org/10.48550/arXiv.2312.00997} {10.48550/arXiv.2312.00997}
  (\bibinfo {year} {2023})\BibitemShut {NoStop}%
\bibitem [{\citenamefont {B\"{a}umer}\ \emph {et~al.}(2023)\citenamefont
  {B\"{a}umer}, \citenamefont {Tripathi}, \citenamefont {Wang}, \citenamefont
  {Rall}, \citenamefont {Chen}, \citenamefont {Majumdar}, \citenamefont
  {Seif},\ and\ \citenamefont {Minev}}]{baumer:2023}%
  \BibitemOpen
  \bibfield  {author} {\bibinfo {author} {\bibfnamefont {E.}~\bibnamefont
  {B\"{a}umer}}, \bibinfo {author} {\bibfnamefont {V.}~\bibnamefont
  {Tripathi}}, \bibinfo {author} {\bibfnamefont {D.~S.}\ \bibnamefont {Wang}},
  \bibinfo {author} {\bibfnamefont {P.}~\bibnamefont {Rall}}, \bibinfo {author}
  {\bibfnamefont {E.~H.}\ \bibnamefont {Chen}}, \bibinfo {author}
  {\bibfnamefont {S.}~\bibnamefont {Majumdar}}, \bibinfo {author}
  {\bibfnamefont {A.}~\bibnamefont {Seif}},\ and\ \bibinfo {author}
  {\bibfnamefont {Z.~K.}\ \bibnamefont {Minev}},\ }\bibfield  {title} {\bibinfo
  {title} {{E}fficient {L}ong-{R}ange {E}ntanglement using {D}ynamic
  {C}ircuits},\ }\bibfield  {journal} {\bibinfo  {journal} {arXiv:2308.13065}\
  }\href {https://doi.org/10.48550/arXiv.2308.13065}
  {10.48550/arXiv.2308.13065} (\bibinfo {year} {2023})\BibitemShut {NoStop}%
\bibitem [{\citenamefont {Acharya}\ \emph {et~al.}(2024)\citenamefont
  {Acharya}, \citenamefont {Abanin}, \citenamefont {Aghababaie-Beni},
  \citenamefont {Aleiner}, \citenamefont {Andersen}, \citenamefont {Ansmann},
  \citenamefont {Arute}, \citenamefont {Arya}, \citenamefont {Asfaw},
  \citenamefont {Astrakhantsev}, \citenamefont {Atalaya}, \citenamefont
  {Babbush}, \citenamefont {Bacon}, \citenamefont {Ballard}, \citenamefont
  {Bardin}, \citenamefont {Bausch}, \citenamefont {Bengtsson}, \citenamefont
  {Bilmes}, \citenamefont {Blackwell}, \citenamefont {Boixo}, \citenamefont
  {Bortoli}, \citenamefont {Bourassa}, \citenamefont {Bovaird}, \citenamefont
  {Brill}, \citenamefont {Broughton}, \citenamefont {Browne}, \citenamefont
  {Buchea}, \citenamefont {Buckley}, \citenamefont {Buell}, \citenamefont
  {Burger}, \citenamefont {Burkett}, \citenamefont {Bushnell}, \citenamefont
  {Cabrera}, \citenamefont {Campero}, \citenamefont {Chang}, \citenamefont
  {Chen}, \citenamefont {Chen}, \citenamefont {Chiaro}, \citenamefont {Chik},
  \citenamefont {Chou}, \citenamefont {Claes}, \citenamefont {Cleland},
  \citenamefont {Cogan}, \citenamefont {Collins}, \citenamefont {Conner},
  \citenamefont {Courtney}, \citenamefont {Crook}, \citenamefont {Curtin},
  \citenamefont {Das}, \citenamefont {Davies}, \citenamefont {De~Lorenzo},
  \citenamefont {Debroy}, \citenamefont {Demura}, \citenamefont {Devoret},
  \citenamefont {Di~Paolo}, \citenamefont {Donohoe}, \citenamefont {Drozdov},
  \citenamefont {Dunsworth}, \citenamefont {Earle}, \citenamefont {Edlich},
  \citenamefont {Eickbusch}, \citenamefont {Elbag}, \citenamefont {Elzouka},
  \citenamefont {Erickson}, \citenamefont {Faoro}, \citenamefont {Farhi},
  \citenamefont {Ferreira}, \citenamefont {Burgos}, \citenamefont {Forati},
  \citenamefont {Fowler}, \citenamefont {Foxen}, \citenamefont {Ganjam},
  \citenamefont {Garcia}, \citenamefont {Gasca}, \citenamefont {Genois},
  \citenamefont {Giang}, \citenamefont {Gidney}, \citenamefont {Gilboa},
  \citenamefont {Gosula}, \citenamefont {Dau}, \citenamefont {Graumann},
  \citenamefont {Greene}, \citenamefont {Gross}, \citenamefont {Habegger},
  \citenamefont {Hall}, \citenamefont {Hamilton}, \citenamefont {Hansen},
  \citenamefont {Harrigan}, \citenamefont {Harrington}, \citenamefont {Heras},
  \citenamefont {Heslin}, \citenamefont {Heu}, \citenamefont {Higgott},
  \citenamefont {Hill}, \citenamefont {Hilton}, \citenamefont {Holland},
  \citenamefont {Hong}, \citenamefont {Huang}, \citenamefont {Huff},
  \citenamefont {Huggins}, \citenamefont {Ioffe}, \citenamefont {Isakov},
  \citenamefont {Iveland}, \citenamefont {Jeffrey}, \citenamefont {Jiang},
  \citenamefont {Jones}, \citenamefont {Jordan}, \citenamefont {Joshi},
  \citenamefont {Juhas}, \citenamefont {Kafri}, \citenamefont {Kang},
  \citenamefont {Karamlou}, \citenamefont {Kechedzhi}, \citenamefont {Kelly},
  \citenamefont {Khaire}, \citenamefont {Khattar}, \citenamefont {Khezri},
  \citenamefont {Kim}, \citenamefont {Klimov}, \citenamefont {Klots},
  \citenamefont {Kobrin}, \citenamefont {Kohli}, \citenamefont {Korotkov},
  \citenamefont {Kostritsa}, \citenamefont {Kothari}, \citenamefont
  {Kozlovskii}, \citenamefont {Kreikebaum}, \citenamefont {Kurilovich},
  \citenamefont {Lacroix}, \citenamefont {Landhuis}, \citenamefont {Lange-Dei},
  \citenamefont {Langley}, \citenamefont {Laptev}, \citenamefont {Lau},
  \citenamefont {Le~Guevel}, \citenamefont {Ledford}, \citenamefont {Lee},
  \citenamefont {Lee}, \citenamefont {Lensky}, \citenamefont {Leon},
  \citenamefont {Lester}, \citenamefont {Li}, \citenamefont {Li}, \citenamefont
  {Lill}, \citenamefont {Liu}, \citenamefont {Livingston}, \citenamefont
  {Locharla}, \citenamefont {Lucero}, \citenamefont {Lundahl}, \citenamefont
  {Lunt}, \citenamefont {Madhuk}, \citenamefont {Malone}, \citenamefont
  {Maloney}, \citenamefont {Mandr{\`a}}, \citenamefont {Manyika}, \citenamefont
  {Martin}, \citenamefont {Martin}, \citenamefont {Martin}, \citenamefont
  {Maxfield}, \citenamefont {McClean}, \citenamefont {McEwen}, \citenamefont
  {Meeks}, \citenamefont {Megrant}, \citenamefont {Mi}, \citenamefont {Miao},
  \citenamefont {Mieszala}, \citenamefont {Molavi}, \citenamefont {Molina},
  \citenamefont {Montazeri}, \citenamefont {Morvan}, \citenamefont {Movassagh},
  \citenamefont {Mruczkiewicz}, \citenamefont {Naaman}, \citenamefont {Neeley},
  \citenamefont {Neill}, \citenamefont {Nersisyan}, \citenamefont {Neven},
  \citenamefont {Newman}, \citenamefont {Ng}, \citenamefont {Nguyen},
  \citenamefont {Nguyen}, \citenamefont {Ni}, \citenamefont {Niu},
  \citenamefont {O'Brien}, \citenamefont {Oliver}, \citenamefont {Opremcak},
  \citenamefont {Ottosson}, \citenamefont {Petukhov}, \citenamefont {Pizzuto},
  \citenamefont {Platt}, \citenamefont {Potter}, \citenamefont {Pritchard},
  \citenamefont {Pryadko}, \citenamefont {Quintana}, \citenamefont
  {Ramachandran}, \citenamefont {Reagor}, \citenamefont {Redding},
  \citenamefont {Rhodes}, \citenamefont {Roberts}, \citenamefont {Rosenberg},
  \citenamefont {Rosenfeld}, \citenamefont {Roushan}, \citenamefont {Rubin},
  \citenamefont {Saei}, \citenamefont {Sank}, \citenamefont {Sankaragomathi},
  \citenamefont {Satzinger}, \citenamefont {Schurkus}, \citenamefont
  {Schuster}, \citenamefont {Senior}, \citenamefont {Shearn}, \citenamefont
  {Shorter}, \citenamefont {Shutty}, \citenamefont {Shvarts}, \citenamefont
  {Singh}, \citenamefont {Sivak}, \citenamefont {Skruzny}, \citenamefont
  {Small}, \citenamefont {Smelyanskiy}, \citenamefont {Smith}, \citenamefont
  {Somma}, \citenamefont {Springer}, \citenamefont {Sterling}, \citenamefont
  {Strain}, \citenamefont {Suchard}, \citenamefont {Szasz}, \citenamefont
  {Sztein}, \citenamefont {Thor}, \citenamefont {Torres}, \citenamefont
  {Torunbalci}, \citenamefont {Vaishnav}, \citenamefont {Vargas}, \citenamefont
  {Vdovichev}, \citenamefont {Vidal}, \citenamefont {Villalonga}, \citenamefont
  {Heidweiller}, \citenamefont {Waltman}, \citenamefont {Wang}, \citenamefont
  {Ware}, \citenamefont {Weber}, \citenamefont {Weidel}, \citenamefont {White},
  \citenamefont {Wong}, \citenamefont {Woo}, \citenamefont {Xing},
  \citenamefont {Yao}, \citenamefont {Yeh}, \citenamefont {Ying}, \citenamefont
  {Yoo}, \citenamefont {Yosri}, \citenamefont {Young}, \citenamefont {Zalcman},
  \citenamefont {Zhang}, \citenamefont {Zhu}, \citenamefont {Zobrist},
  \citenamefont {AI},\ and\ \citenamefont {Collaborators}}]{acharya:2024}%
  \BibitemOpen
  \bibfield  {author} {\bibinfo {author} {\bibfnamefont {R.}~\bibnamefont
  {Acharya}}, \bibinfo {author} {\bibfnamefont {D.~A.}\ \bibnamefont {Abanin}},
  \bibinfo {author} {\bibfnamefont {L.}~\bibnamefont {Aghababaie-Beni}},
  \bibinfo {author} {\bibfnamefont {I.}~\bibnamefont {Aleiner}}, \bibinfo
  {author} {\bibfnamefont {T.~I.}\ \bibnamefont {Andersen}}, \bibinfo {author}
  {\bibfnamefont {M.}~\bibnamefont {Ansmann}}, \bibinfo {author} {\bibfnamefont
  {F.}~\bibnamefont {Arute}}, \bibinfo {author} {\bibfnamefont
  {K.}~\bibnamefont {Arya}}, \bibinfo {author} {\bibfnamefont {A.}~\bibnamefont
  {Asfaw}}, \bibinfo {author} {\bibfnamefont {N.}~\bibnamefont
  {Astrakhantsev}}, \bibinfo {author} {\bibfnamefont {J.}~\bibnamefont
  {Atalaya}}, \bibinfo {author} {\bibfnamefont {R.}~\bibnamefont {Babbush}},
  \bibinfo {author} {\bibfnamefont {D.}~\bibnamefont {Bacon}}, \bibinfo
  {author} {\bibfnamefont {B.}~\bibnamefont {Ballard}}, \bibinfo {author}
  {\bibfnamefont {J.~C.}\ \bibnamefont {Bardin}}, \bibinfo {author}
  {\bibfnamefont {J.}~\bibnamefont {Bausch}}, \bibinfo {author} {\bibfnamefont
  {A.}~\bibnamefont {Bengtsson}}, \bibinfo {author} {\bibfnamefont
  {A.}~\bibnamefont {Bilmes}}, \bibinfo {author} {\bibfnamefont
  {S.}~\bibnamefont {Blackwell}}, \bibinfo {author} {\bibfnamefont
  {S.}~\bibnamefont {Boixo}}, \bibinfo {author} {\bibfnamefont
  {G.}~\bibnamefont {Bortoli}}, \bibinfo {author} {\bibfnamefont
  {A.}~\bibnamefont {Bourassa}}, \bibinfo {author} {\bibfnamefont
  {J.}~\bibnamefont {Bovaird}}, \bibinfo {author} {\bibfnamefont
  {L.}~\bibnamefont {Brill}}, \bibinfo {author} {\bibfnamefont
  {M.}~\bibnamefont {Broughton}}, \bibinfo {author} {\bibfnamefont {D.~A.}\
  \bibnamefont {Browne}}, \bibinfo {author} {\bibfnamefont {B.}~\bibnamefont
  {Buchea}}, \bibinfo {author} {\bibfnamefont {B.~B.}\ \bibnamefont {Buckley}},
  \bibinfo {author} {\bibfnamefont {D.~A.}\ \bibnamefont {Buell}}, \bibinfo
  {author} {\bibfnamefont {T.}~\bibnamefont {Burger}}, \bibinfo {author}
  {\bibfnamefont {B.}~\bibnamefont {Burkett}}, \bibinfo {author} {\bibfnamefont
  {N.}~\bibnamefont {Bushnell}}, \bibinfo {author} {\bibfnamefont
  {A.}~\bibnamefont {Cabrera}}, \bibinfo {author} {\bibfnamefont
  {J.}~\bibnamefont {Campero}}, \bibinfo {author} {\bibfnamefont {H.-S.}\
  \bibnamefont {Chang}}, \bibinfo {author} {\bibfnamefont {Y.}~\bibnamefont
  {Chen}}, \bibinfo {author} {\bibfnamefont {Z.}~\bibnamefont {Chen}}, \bibinfo
  {author} {\bibfnamefont {B.}~\bibnamefont {Chiaro}}, \bibinfo {author}
  {\bibfnamefont {D.}~\bibnamefont {Chik}}, \bibinfo {author} {\bibfnamefont
  {C.}~\bibnamefont {Chou}}, \bibinfo {author} {\bibfnamefont {J.}~\bibnamefont
  {Claes}}, \bibinfo {author} {\bibfnamefont {A.~Y.}\ \bibnamefont {Cleland}},
  \bibinfo {author} {\bibfnamefont {J.}~\bibnamefont {Cogan}}, \bibinfo
  {author} {\bibfnamefont {R.}~\bibnamefont {Collins}}, \bibinfo {author}
  {\bibfnamefont {P.}~\bibnamefont {Conner}}, \bibinfo {author} {\bibfnamefont
  {W.}~\bibnamefont {Courtney}}, \bibinfo {author} {\bibfnamefont {A.~L.}\
  \bibnamefont {Crook}}, \bibinfo {author} {\bibfnamefont {B.}~\bibnamefont
  {Curtin}}, \bibinfo {author} {\bibfnamefont {S.}~\bibnamefont {Das}},
  \bibinfo {author} {\bibfnamefont {A.}~\bibnamefont {Davies}}, \bibinfo
  {author} {\bibfnamefont {L.}~\bibnamefont {De~Lorenzo}}, \bibinfo {author}
  {\bibfnamefont {D.~M.}\ \bibnamefont {Debroy}}, \bibinfo {author}
  {\bibfnamefont {S.}~\bibnamefont {Demura}}, \bibinfo {author} {\bibfnamefont
  {M.}~\bibnamefont {Devoret}}, \bibinfo {author} {\bibfnamefont
  {A.}~\bibnamefont {Di~Paolo}}, \bibinfo {author} {\bibfnamefont
  {P.}~\bibnamefont {Donohoe}}, \bibinfo {author} {\bibfnamefont
  {I.}~\bibnamefont {Drozdov}}, \bibinfo {author} {\bibfnamefont
  {A.}~\bibnamefont {Dunsworth}}, \bibinfo {author} {\bibfnamefont
  {C.}~\bibnamefont {Earle}}, \bibinfo {author} {\bibfnamefont
  {T.}~\bibnamefont {Edlich}}, \bibinfo {author} {\bibfnamefont
  {A.}~\bibnamefont {Eickbusch}}, \bibinfo {author} {\bibfnamefont {A.~M.}\
  \bibnamefont {Elbag}}, \bibinfo {author} {\bibfnamefont {M.}~\bibnamefont
  {Elzouka}}, \bibinfo {author} {\bibfnamefont {C.}~\bibnamefont {Erickson}},
  \bibinfo {author} {\bibfnamefont {L.}~\bibnamefont {Faoro}}, \bibinfo
  {author} {\bibfnamefont {E.}~\bibnamefont {Farhi}}, \bibinfo {author}
  {\bibfnamefont {V.~S.}\ \bibnamefont {Ferreira}}, \bibinfo {author}
  {\bibfnamefont {L.~F.}\ \bibnamefont {Burgos}}, \bibinfo {author}
  {\bibfnamefont {E.}~\bibnamefont {Forati}}, \bibinfo {author} {\bibfnamefont
  {A.~G.}\ \bibnamefont {Fowler}}, \bibinfo {author} {\bibfnamefont
  {B.}~\bibnamefont {Foxen}}, \bibinfo {author} {\bibfnamefont
  {S.}~\bibnamefont {Ganjam}}, \bibinfo {author} {\bibfnamefont
  {G.}~\bibnamefont {Garcia}}, \bibinfo {author} {\bibfnamefont
  {R.}~\bibnamefont {Gasca}}, \bibinfo {author} {\bibfnamefont
  {{\'E}.}~\bibnamefont {Genois}}, \bibinfo {author} {\bibfnamefont
  {W.}~\bibnamefont {Giang}}, \bibinfo {author} {\bibfnamefont
  {C.}~\bibnamefont {Gidney}}, \bibinfo {author} {\bibfnamefont
  {D.}~\bibnamefont {Gilboa}}, \bibinfo {author} {\bibfnamefont
  {R.}~\bibnamefont {Gosula}}, \bibinfo {author} {\bibfnamefont {A.~G.}\
  \bibnamefont {Dau}}, \bibinfo {author} {\bibfnamefont {D.}~\bibnamefont
  {Graumann}}, \bibinfo {author} {\bibfnamefont {A.}~\bibnamefont {Greene}},
  \bibinfo {author} {\bibfnamefont {J.~A.}\ \bibnamefont {Gross}}, \bibinfo
  {author} {\bibfnamefont {S.}~\bibnamefont {Habegger}}, \bibinfo {author}
  {\bibfnamefont {J.}~\bibnamefont {Hall}}, \bibinfo {author} {\bibfnamefont
  {M.~C.}\ \bibnamefont {Hamilton}}, \bibinfo {author} {\bibfnamefont
  {M.}~\bibnamefont {Hansen}}, \bibinfo {author} {\bibfnamefont {M.~P.}\
  \bibnamefont {Harrigan}}, \bibinfo {author} {\bibfnamefont {S.~D.}\
  \bibnamefont {Harrington}}, \bibinfo {author} {\bibfnamefont {F.~J.~H.}\
  \bibnamefont {Heras}}, \bibinfo {author} {\bibfnamefont {S.}~\bibnamefont
  {Heslin}}, \bibinfo {author} {\bibfnamefont {P.}~\bibnamefont {Heu}},
  \bibinfo {author} {\bibfnamefont {O.}~\bibnamefont {Higgott}}, \bibinfo
  {author} {\bibfnamefont {G.}~\bibnamefont {Hill}}, \bibinfo {author}
  {\bibfnamefont {J.}~\bibnamefont {Hilton}}, \bibinfo {author} {\bibfnamefont
  {G.}~\bibnamefont {Holland}}, \bibinfo {author} {\bibfnamefont
  {S.}~\bibnamefont {Hong}}, \bibinfo {author} {\bibfnamefont {H.-Y.}\
  \bibnamefont {Huang}}, \bibinfo {author} {\bibfnamefont {A.}~\bibnamefont
  {Huff}}, \bibinfo {author} {\bibfnamefont {W.~J.}\ \bibnamefont {Huggins}},
  \bibinfo {author} {\bibfnamefont {L.~B.}\ \bibnamefont {Ioffe}}, \bibinfo
  {author} {\bibfnamefont {S.~V.}\ \bibnamefont {Isakov}}, \bibinfo {author}
  {\bibfnamefont {J.}~\bibnamefont {Iveland}}, \bibinfo {author} {\bibfnamefont
  {E.}~\bibnamefont {Jeffrey}}, \bibinfo {author} {\bibfnamefont
  {Z.}~\bibnamefont {Jiang}}, \bibinfo {author} {\bibfnamefont
  {C.}~\bibnamefont {Jones}}, \bibinfo {author} {\bibfnamefont
  {S.}~\bibnamefont {Jordan}}, \bibinfo {author} {\bibfnamefont
  {C.}~\bibnamefont {Joshi}}, \bibinfo {author} {\bibfnamefont
  {P.}~\bibnamefont {Juhas}}, \bibinfo {author} {\bibfnamefont
  {D.}~\bibnamefont {Kafri}}, \bibinfo {author} {\bibfnamefont
  {H.}~\bibnamefont {Kang}}, \bibinfo {author} {\bibfnamefont {A.~H.}\
  \bibnamefont {Karamlou}}, \bibinfo {author} {\bibfnamefont {K.}~\bibnamefont
  {Kechedzhi}}, \bibinfo {author} {\bibfnamefont {J.}~\bibnamefont {Kelly}},
  \bibinfo {author} {\bibfnamefont {T.}~\bibnamefont {Khaire}}, \bibinfo
  {author} {\bibfnamefont {T.}~\bibnamefont {Khattar}}, \bibinfo {author}
  {\bibfnamefont {M.}~\bibnamefont {Khezri}}, \bibinfo {author} {\bibfnamefont
  {S.}~\bibnamefont {Kim}}, \bibinfo {author} {\bibfnamefont {P.~V.}\
  \bibnamefont {Klimov}}, \bibinfo {author} {\bibfnamefont {A.~R.}\
  \bibnamefont {Klots}}, \bibinfo {author} {\bibfnamefont {B.}~\bibnamefont
  {Kobrin}}, \bibinfo {author} {\bibfnamefont {P.}~\bibnamefont {Kohli}},
  \bibinfo {author} {\bibfnamefont {A.~N.}\ \bibnamefont {Korotkov}}, \bibinfo
  {author} {\bibfnamefont {F.}~\bibnamefont {Kostritsa}}, \bibinfo {author}
  {\bibfnamefont {R.}~\bibnamefont {Kothari}}, \bibinfo {author} {\bibfnamefont
  {B.}~\bibnamefont {Kozlovskii}}, \bibinfo {author} {\bibfnamefont {J.~M.}\
  \bibnamefont {Kreikebaum}}, \bibinfo {author} {\bibfnamefont {V.~D.}\
  \bibnamefont {Kurilovich}}, \bibinfo {author} {\bibfnamefont
  {N.}~\bibnamefont {Lacroix}}, \bibinfo {author} {\bibfnamefont
  {D.}~\bibnamefont {Landhuis}}, \bibinfo {author} {\bibfnamefont
  {T.}~\bibnamefont {Lange-Dei}}, \bibinfo {author} {\bibfnamefont {B.~W.}\
  \bibnamefont {Langley}}, \bibinfo {author} {\bibfnamefont {P.}~\bibnamefont
  {Laptev}}, \bibinfo {author} {\bibfnamefont {K.-M.}\ \bibnamefont {Lau}},
  \bibinfo {author} {\bibfnamefont {L.}~\bibnamefont {Le~Guevel}}, \bibinfo
  {author} {\bibfnamefont {J.}~\bibnamefont {Ledford}}, \bibinfo {author}
  {\bibfnamefont {J.}~\bibnamefont {Lee}}, \bibinfo {author} {\bibfnamefont
  {K.}~\bibnamefont {Lee}}, \bibinfo {author} {\bibfnamefont {Y.~D.}\
  \bibnamefont {Lensky}}, \bibinfo {author} {\bibfnamefont {S.}~\bibnamefont
  {Leon}}, \bibinfo {author} {\bibfnamefont {B.~J.}\ \bibnamefont {Lester}},
  \bibinfo {author} {\bibfnamefont {W.~Y.}\ \bibnamefont {Li}}, \bibinfo
  {author} {\bibfnamefont {Y.}~\bibnamefont {Li}}, \bibinfo {author}
  {\bibfnamefont {A.~T.}\ \bibnamefont {Lill}}, \bibinfo {author}
  {\bibfnamefont {W.}~\bibnamefont {Liu}}, \bibinfo {author} {\bibfnamefont
  {W.~P.}\ \bibnamefont {Livingston}}, \bibinfo {author} {\bibfnamefont
  {A.}~\bibnamefont {Locharla}}, \bibinfo {author} {\bibfnamefont
  {E.}~\bibnamefont {Lucero}}, \bibinfo {author} {\bibfnamefont
  {D.}~\bibnamefont {Lundahl}}, \bibinfo {author} {\bibfnamefont
  {A.}~\bibnamefont {Lunt}}, \bibinfo {author} {\bibfnamefont {S.}~\bibnamefont
  {Madhuk}}, \bibinfo {author} {\bibfnamefont {F.~D.}\ \bibnamefont {Malone}},
  \bibinfo {author} {\bibfnamefont {A.}~\bibnamefont {Maloney}}, \bibinfo
  {author} {\bibfnamefont {S.}~\bibnamefont {Mandr{\`a}}}, \bibinfo {author}
  {\bibfnamefont {J.}~\bibnamefont {Manyika}}, \bibinfo {author} {\bibfnamefont
  {L.~S.}\ \bibnamefont {Martin}}, \bibinfo {author} {\bibfnamefont
  {O.}~\bibnamefont {Martin}}, \bibinfo {author} {\bibfnamefont
  {S.}~\bibnamefont {Martin}}, \bibinfo {author} {\bibfnamefont
  {C.}~\bibnamefont {Maxfield}}, \bibinfo {author} {\bibfnamefont {J.~R.}\
  \bibnamefont {McClean}}, \bibinfo {author} {\bibfnamefont {M.}~\bibnamefont
  {McEwen}}, \bibinfo {author} {\bibfnamefont {S.}~\bibnamefont {Meeks}},
  \bibinfo {author} {\bibfnamefont {A.}~\bibnamefont {Megrant}}, \bibinfo
  {author} {\bibfnamefont {X.}~\bibnamefont {Mi}}, \bibinfo {author}
  {\bibfnamefont {K.~C.}\ \bibnamefont {Miao}}, \bibinfo {author}
  {\bibfnamefont {A.}~\bibnamefont {Mieszala}}, \bibinfo {author}
  {\bibfnamefont {R.}~\bibnamefont {Molavi}}, \bibinfo {author} {\bibfnamefont
  {S.}~\bibnamefont {Molina}}, \bibinfo {author} {\bibfnamefont
  {S.}~\bibnamefont {Montazeri}}, \bibinfo {author} {\bibfnamefont
  {A.}~\bibnamefont {Morvan}}, \bibinfo {author} {\bibfnamefont
  {R.}~\bibnamefont {Movassagh}}, \bibinfo {author} {\bibfnamefont
  {W.}~\bibnamefont {Mruczkiewicz}}, \bibinfo {author} {\bibfnamefont
  {O.}~\bibnamefont {Naaman}}, \bibinfo {author} {\bibfnamefont
  {M.}~\bibnamefont {Neeley}}, \bibinfo {author} {\bibfnamefont
  {C.}~\bibnamefont {Neill}}, \bibinfo {author} {\bibfnamefont
  {A.}~\bibnamefont {Nersisyan}}, \bibinfo {author} {\bibfnamefont
  {H.}~\bibnamefont {Neven}}, \bibinfo {author} {\bibfnamefont
  {M.}~\bibnamefont {Newman}}, \bibinfo {author} {\bibfnamefont {J.~H.}\
  \bibnamefont {Ng}}, \bibinfo {author} {\bibfnamefont {A.}~\bibnamefont
  {Nguyen}}, \bibinfo {author} {\bibfnamefont {M.}~\bibnamefont {Nguyen}},
  \bibinfo {author} {\bibfnamefont {C.-H.}\ \bibnamefont {Ni}}, \bibinfo
  {author} {\bibfnamefont {M.~Y.}\ \bibnamefont {Niu}}, \bibinfo {author}
  {\bibfnamefont {T.~E.}\ \bibnamefont {O'Brien}}, \bibinfo {author}
  {\bibfnamefont {W.~D.}\ \bibnamefont {Oliver}}, \bibinfo {author}
  {\bibfnamefont {A.}~\bibnamefont {Opremcak}}, \bibinfo {author}
  {\bibfnamefont {K.}~\bibnamefont {Ottosson}}, \bibinfo {author}
  {\bibfnamefont {A.}~\bibnamefont {Petukhov}}, \bibinfo {author}
  {\bibfnamefont {A.}~\bibnamefont {Pizzuto}}, \bibinfo {author} {\bibfnamefont
  {J.}~\bibnamefont {Platt}}, \bibinfo {author} {\bibfnamefont
  {R.}~\bibnamefont {Potter}}, \bibinfo {author} {\bibfnamefont
  {O.}~\bibnamefont {Pritchard}}, \bibinfo {author} {\bibfnamefont {L.~P.}\
  \bibnamefont {Pryadko}}, \bibinfo {author} {\bibfnamefont {C.}~\bibnamefont
  {Quintana}}, \bibinfo {author} {\bibfnamefont {G.}~\bibnamefont
  {Ramachandran}}, \bibinfo {author} {\bibfnamefont {M.~J.}\ \bibnamefont
  {Reagor}}, \bibinfo {author} {\bibfnamefont {J.}~\bibnamefont {Redding}},
  \bibinfo {author} {\bibfnamefont {D.~M.}\ \bibnamefont {Rhodes}}, \bibinfo
  {author} {\bibfnamefont {G.}~\bibnamefont {Roberts}}, \bibinfo {author}
  {\bibfnamefont {E.}~\bibnamefont {Rosenberg}}, \bibinfo {author}
  {\bibfnamefont {E.}~\bibnamefont {Rosenfeld}}, \bibinfo {author}
  {\bibfnamefont {P.}~\bibnamefont {Roushan}}, \bibinfo {author} {\bibfnamefont
  {N.~C.}\ \bibnamefont {Rubin}}, \bibinfo {author} {\bibfnamefont
  {N.}~\bibnamefont {Saei}}, \bibinfo {author} {\bibfnamefont {D.}~\bibnamefont
  {Sank}}, \bibinfo {author} {\bibfnamefont {K.}~\bibnamefont
  {Sankaragomathi}}, \bibinfo {author} {\bibfnamefont {K.~J.}\ \bibnamefont
  {Satzinger}}, \bibinfo {author} {\bibfnamefont {H.~F.}\ \bibnamefont
  {Schurkus}}, \bibinfo {author} {\bibfnamefont {C.}~\bibnamefont {Schuster}},
  \bibinfo {author} {\bibfnamefont {A.~W.}\ \bibnamefont {Senior}}, \bibinfo
  {author} {\bibfnamefont {M.~J.}\ \bibnamefont {Shearn}}, \bibinfo {author}
  {\bibfnamefont {A.}~\bibnamefont {Shorter}}, \bibinfo {author} {\bibfnamefont
  {N.}~\bibnamefont {Shutty}}, \bibinfo {author} {\bibfnamefont
  {V.}~\bibnamefont {Shvarts}}, \bibinfo {author} {\bibfnamefont
  {S.}~\bibnamefont {Singh}}, \bibinfo {author} {\bibfnamefont
  {V.}~\bibnamefont {Sivak}}, \bibinfo {author} {\bibfnamefont
  {J.}~\bibnamefont {Skruzny}}, \bibinfo {author} {\bibfnamefont
  {S.}~\bibnamefont {Small}}, \bibinfo {author} {\bibfnamefont
  {V.}~\bibnamefont {Smelyanskiy}}, \bibinfo {author} {\bibfnamefont {W.~C.}\
  \bibnamefont {Smith}}, \bibinfo {author} {\bibfnamefont {R.~D.}\ \bibnamefont
  {Somma}}, \bibinfo {author} {\bibfnamefont {S.}~\bibnamefont {Springer}},
  \bibinfo {author} {\bibfnamefont {G.}~\bibnamefont {Sterling}}, \bibinfo
  {author} {\bibfnamefont {D.}~\bibnamefont {Strain}}, \bibinfo {author}
  {\bibfnamefont {J.}~\bibnamefont {Suchard}}, \bibinfo {author} {\bibfnamefont
  {A.}~\bibnamefont {Szasz}}, \bibinfo {author} {\bibfnamefont
  {A.}~\bibnamefont {Sztein}}, \bibinfo {author} {\bibfnamefont
  {D.}~\bibnamefont {Thor}}, \bibinfo {author} {\bibfnamefont {A.}~\bibnamefont
  {Torres}}, \bibinfo {author} {\bibfnamefont {M.~M.}\ \bibnamefont
  {Torunbalci}}, \bibinfo {author} {\bibfnamefont {A.}~\bibnamefont
  {Vaishnav}}, \bibinfo {author} {\bibfnamefont {J.}~\bibnamefont {Vargas}},
  \bibinfo {author} {\bibfnamefont {S.}~\bibnamefont {Vdovichev}}, \bibinfo
  {author} {\bibfnamefont {G.}~\bibnamefont {Vidal}}, \bibinfo {author}
  {\bibfnamefont {B.}~\bibnamefont {Villalonga}}, \bibinfo {author}
  {\bibfnamefont {C.~V.}\ \bibnamefont {Heidweiller}}, \bibinfo {author}
  {\bibfnamefont {S.}~\bibnamefont {Waltman}}, \bibinfo {author} {\bibfnamefont
  {S.~X.}\ \bibnamefont {Wang}}, \bibinfo {author} {\bibfnamefont
  {B.}~\bibnamefont {Ware}}, \bibinfo {author} {\bibfnamefont {K.}~\bibnamefont
  {Weber}}, \bibinfo {author} {\bibfnamefont {T.}~\bibnamefont {Weidel}},
  \bibinfo {author} {\bibfnamefont {T.}~\bibnamefont {White}}, \bibinfo
  {author} {\bibfnamefont {K.}~\bibnamefont {Wong}}, \bibinfo {author}
  {\bibfnamefont {B.~W.~K.}\ \bibnamefont {Woo}}, \bibinfo {author}
  {\bibfnamefont {C.}~\bibnamefont {Xing}}, \bibinfo {author} {\bibfnamefont
  {Z.~J.}\ \bibnamefont {Yao}}, \bibinfo {author} {\bibfnamefont
  {P.}~\bibnamefont {Yeh}}, \bibinfo {author} {\bibfnamefont {B.}~\bibnamefont
  {Ying}}, \bibinfo {author} {\bibfnamefont {J.}~\bibnamefont {Yoo}}, \bibinfo
  {author} {\bibfnamefont {N.}~\bibnamefont {Yosri}}, \bibinfo {author}
  {\bibfnamefont {G.}~\bibnamefont {Young}}, \bibinfo {author} {\bibfnamefont
  {A.}~\bibnamefont {Zalcman}}, \bibinfo {author} {\bibfnamefont
  {Y.}~\bibnamefont {Zhang}}, \bibinfo {author} {\bibfnamefont
  {N.}~\bibnamefont {Zhu}}, \bibinfo {author} {\bibfnamefont {N.}~\bibnamefont
  {Zobrist}}, \bibinfo {author} {\bibfnamefont {G.~Q.}\ \bibnamefont {AI}},\
  and\ \bibinfo {author} {\bibnamefont {Collaborators}},\ }\bibfield  {title}
  {\bibinfo {title} {Quantum error correction below the surface code
  threshold},\ }\bibfield  {journal} {\bibinfo  {journal} {Nature}\ }\href
  {https://doi.org/10.1038/s41586-024-08449-y} {10.1038/s41586-024-08449-y}
  (\bibinfo {year} {2024})\BibitemShut {NoStop}%
\bibitem [{\citenamefont {Bluvstein}\ \emph {et~al.}(2024)\citenamefont
  {Bluvstein}, \citenamefont {Evered}, \citenamefont {Geim}, \citenamefont
  {Li}, \citenamefont {Zhou}, \citenamefont {Manovitz}, \citenamefont {Ebadi},
  \citenamefont {Cain}, \citenamefont {Kalinowski}, \citenamefont {Hangleiter},
  \citenamefont {Bonilla~Ataides}, \citenamefont {Maskara}, \citenamefont
  {Cong}, \citenamefont {Gao}, \citenamefont {Sales~Rodriguez}, \citenamefont
  {Karolyshyn}, \citenamefont {Semeghini}, \citenamefont {Gullans},
  \citenamefont {Greiner}, \citenamefont {Vuleti{\'c}},\ and\ \citenamefont
  {Lukin}}]{bluvstein:2024}%
  \BibitemOpen
  \bibfield  {author} {\bibinfo {author} {\bibfnamefont {D.}~\bibnamefont
  {Bluvstein}}, \bibinfo {author} {\bibfnamefont {S.~J.}\ \bibnamefont
  {Evered}}, \bibinfo {author} {\bibfnamefont {A.~A.}\ \bibnamefont {Geim}},
  \bibinfo {author} {\bibfnamefont {S.~H.}\ \bibnamefont {Li}}, \bibinfo
  {author} {\bibfnamefont {H.}~\bibnamefont {Zhou}}, \bibinfo {author}
  {\bibfnamefont {T.}~\bibnamefont {Manovitz}}, \bibinfo {author}
  {\bibfnamefont {S.}~\bibnamefont {Ebadi}}, \bibinfo {author} {\bibfnamefont
  {M.}~\bibnamefont {Cain}}, \bibinfo {author} {\bibfnamefont {M.}~\bibnamefont
  {Kalinowski}}, \bibinfo {author} {\bibfnamefont {D.}~\bibnamefont
  {Hangleiter}}, \bibinfo {author} {\bibfnamefont {J.~P.}\ \bibnamefont
  {Bonilla~Ataides}}, \bibinfo {author} {\bibfnamefont {N.}~\bibnamefont
  {Maskara}}, \bibinfo {author} {\bibfnamefont {I.}~\bibnamefont {Cong}},
  \bibinfo {author} {\bibfnamefont {X.}~\bibnamefont {Gao}}, \bibinfo {author}
  {\bibfnamefont {P.}~\bibnamefont {Sales~Rodriguez}}, \bibinfo {author}
  {\bibfnamefont {T.}~\bibnamefont {Karolyshyn}}, \bibinfo {author}
  {\bibfnamefont {G.}~\bibnamefont {Semeghini}}, \bibinfo {author}
  {\bibfnamefont {M.~J.}\ \bibnamefont {Gullans}}, \bibinfo {author}
  {\bibfnamefont {M.}~\bibnamefont {Greiner}}, \bibinfo {author} {\bibfnamefont
  {V.}~\bibnamefont {Vuleti{\'c}}},\ and\ \bibinfo {author} {\bibfnamefont
  {M.~D.}\ \bibnamefont {Lukin}},\ }\bibfield  {title} {\bibinfo {title}
  {Logical quantum processor based on reconfigurable atom arrays},\ }\href
  {https://doi.org/10.1038/s41586-023-06927-3} {\bibfield  {journal} {\bibinfo
  {journal} {Nature}\ }\textbf {\bibinfo {volume} {626}},\ \bibinfo {pages}
  {58} (\bibinfo {year} {2024})}\BibitemShut {NoStop}%
\bibitem [{\citenamefont {Farrell}\ \emph
  {et~al.}(2024{\natexlab{b}})\citenamefont {Farrell}, \citenamefont {Illa},
  \citenamefont {Ciavarella},\ and\ \citenamefont {Savage}}]{farrell:2024b}%
  \BibitemOpen
  \bibfield  {author} {\bibinfo {author} {\bibfnamefont {R.~C.}\ \bibnamefont
  {Farrell}}, \bibinfo {author} {\bibfnamefont {M.}~\bibnamefont {Illa}},
  \bibinfo {author} {\bibfnamefont {A.~N.}\ \bibnamefont {Ciavarella}},\ and\
  \bibinfo {author} {\bibfnamefont {M.~J.}\ \bibnamefont {Savage}},\ }\bibfield
   {title} {\bibinfo {title} {Quantum simulations of hadron dynamics in the
  schwinger model using 112 qubits},\ }\href
  {https://doi.org/10.1103/PhysRevD.109.114510} {\bibfield  {journal} {\bibinfo
   {journal} {Phys. Rev. D}\ }\textbf {\bibinfo {volume} {109}},\ \bibinfo
  {pages} {114510} (\bibinfo {year} {2024}{\natexlab{b}})}\BibitemShut
  {NoStop}%
\bibitem [{\citenamefont {Shinjo}\ \emph {et~al.}(2024)\citenamefont {Shinjo},
  \citenamefont {Seki}, \citenamefont {Shirakawa}, \citenamefont {Sun},\ and\
  \citenamefont {Yunoki}}]{shinjo:2024}%
  \BibitemOpen
  \bibfield  {author} {\bibinfo {author} {\bibfnamefont {K.}~\bibnamefont
  {Shinjo}}, \bibinfo {author} {\bibfnamefont {K.}~\bibnamefont {Seki}},
  \bibinfo {author} {\bibfnamefont {T.}~\bibnamefont {Shirakawa}}, \bibinfo
  {author} {\bibfnamefont {R.-Y.}\ \bibnamefont {Sun}},\ and\ \bibinfo {author}
  {\bibfnamefont {S.}~\bibnamefont {Yunoki}},\ }\bibfield  {title} {\bibinfo
  {title} {{U}nveiling clean two-dimensional discrete time quasicrystals on a
  digital quantum computer},\ }\bibfield  {journal} {\bibinfo  {journal}
  {arXiv:2403.16718}\ }\href {https://doi.org/10.48550/arXiv.2403.16718}
  {10.48550/arXiv.2403.16718} (\bibinfo {year} {2024})\BibitemShut {NoStop}%
\bibitem [{\citenamefont {Miessen}\ \emph {et~al.}(2024)\citenamefont
  {Miessen}, \citenamefont {Egger}, \citenamefont {Tavernelli},\ and\
  \citenamefont {Mazzola}}]{miessen:2024}%
  \BibitemOpen
  \bibfield  {author} {\bibinfo {author} {\bibfnamefont {A.}~\bibnamefont
  {Miessen}}, \bibinfo {author} {\bibfnamefont {D.~J.}\ \bibnamefont {Egger}},
  \bibinfo {author} {\bibfnamefont {I.}~\bibnamefont {Tavernelli}},\ and\
  \bibinfo {author} {\bibfnamefont {G.}~\bibnamefont {Mazzola}},\ }\bibfield
  {title} {\bibinfo {title} {{B}enchmarking digital quantum simulations above
  hundreds of qubits using quantum critical dynamics},\ }\bibfield  {journal}
  {\bibinfo  {journal} {arXiv:2404.08053}\ }\href
  {https://doi.org/10.48550/arXiv.2404.08053} {10.48550/arXiv.2404.08053}
  (\bibinfo {year} {2024})\BibitemShut {NoStop}%
\bibitem [{\citenamefont {Robledo-Moreno}\ \emph {et~al.}(2024)\citenamefont
  {Robledo-Moreno}, \citenamefont {Motta}, \citenamefont {Haas}, \citenamefont
  {Javadi-Abhari}, \citenamefont {Jurcevic}, \citenamefont {William},
  \citenamefont {Martiel}, \citenamefont {Sharma}, \citenamefont {Sharma},
  \citenamefont {Shirakawa}, \citenamefont {Sitdikov}, \citenamefont {Sun},
  \citenamefont {Sung}, \citenamefont {Takita}, \citenamefont {Tran},
  \citenamefont {Yunoki},\ and\ \citenamefont {Mezzacapo}}]{robledo:2024}%
  \BibitemOpen
  \bibfield  {author} {\bibinfo {author} {\bibfnamefont {J.}~\bibnamefont
  {Robledo-Moreno}}, \bibinfo {author} {\bibfnamefont {M.}~\bibnamefont
  {Motta}}, \bibinfo {author} {\bibfnamefont {H.}~\bibnamefont {Haas}},
  \bibinfo {author} {\bibfnamefont {A.}~\bibnamefont {Javadi-Abhari}}, \bibinfo
  {author} {\bibfnamefont {P.}~\bibnamefont {Jurcevic}}, \bibinfo {author}
  {\bibfnamefont {K.}~\bibnamefont {William}}, \bibinfo {author} {\bibfnamefont
  {S.}~\bibnamefont {Martiel}}, \bibinfo {author} {\bibfnamefont
  {K.}~\bibnamefont {Sharma}}, \bibinfo {author} {\bibfnamefont
  {S.}~\bibnamefont {Sharma}}, \bibinfo {author} {\bibfnamefont
  {T.}~\bibnamefont {Shirakawa}}, \bibinfo {author} {\bibfnamefont
  {I.}~\bibnamefont {Sitdikov}}, \bibinfo {author} {\bibfnamefont {R.-Y.}\
  \bibnamefont {Sun}}, \bibinfo {author} {\bibfnamefont {K.~J.}\ \bibnamefont
  {Sung}}, \bibinfo {author} {\bibfnamefont {M.}~\bibnamefont {Takita}},
  \bibinfo {author} {\bibfnamefont {M.~C.}\ \bibnamefont {Tran}}, \bibinfo
  {author} {\bibfnamefont {S.}~\bibnamefont {Yunoki}},\ and\ \bibinfo {author}
  {\bibfnamefont {A.}~\bibnamefont {Mezzacapo}},\ }\bibfield  {title} {\bibinfo
  {title} {{C}hemistry {B}eyond {E}xact {S}olutions on a {Q}uantum-{C}entric
  {S}upercomputer},\ }\bibfield  {journal} {\bibinfo  {journal}
  {arXiv:2405.05068}\ }\href {https://doi.org/10.48550/arXiv.2405.05068}
  {10.48550/arXiv.2405.05068} (\bibinfo {year} {2024})\BibitemShut {NoStop}%
\bibitem [{\citenamefont {Montanez-Barrera}\ and\ \citenamefont
  {Michielsen}(2024)}]{montanez:2024}%
  \BibitemOpen
  \bibfield  {author} {\bibinfo {author} {\bibfnamefont {J.~A.}\ \bibnamefont
  {Montanez-Barrera}}\ and\ \bibinfo {author} {\bibfnamefont {K.}~\bibnamefont
  {Michielsen}},\ }\bibfield  {title} {\bibinfo {title} {{T}owards a universal
  {QAOA} protocol: {E}vidence of a scaling advantage in solving some
  combinatorial optimization problems},\ }\bibfield  {journal} {\bibinfo
  {journal} {arXiv:2405.09169}\ }\href
  {https://doi.org/10.48550/arXiv.2405.09169} {10.48550/arXiv.2405.09169}
  (\bibinfo {year} {2024})\BibitemShut {NoStop}%
\bibitem [{\citenamefont {Cadavid}\ \emph {et~al.}(2024)\citenamefont
  {Cadavid}, \citenamefont {Dalal}, \citenamefont {Simen}, \citenamefont
  {Solano},\ and\ \citenamefont {Hegade}}]{cadavid:2024}%
  \BibitemOpen
  \bibfield  {author} {\bibinfo {author} {\bibfnamefont {A.~G.}\ \bibnamefont
  {Cadavid}}, \bibinfo {author} {\bibfnamefont {A.}~\bibnamefont {Dalal}},
  \bibinfo {author} {\bibfnamefont {A.}~\bibnamefont {Simen}}, \bibinfo
  {author} {\bibfnamefont {E.}~\bibnamefont {Solano}},\ and\ \bibinfo {author}
  {\bibfnamefont {N.~N.}\ \bibnamefont {Hegade}},\ }\bibfield  {title}
  {\bibinfo {title} {{B}ias-field digitized counterdiabatic quantum
  optimization},\ }\bibfield  {journal} {\bibinfo  {journal}
  {arXiv:2405.13898}\ }\href {https://doi.org/10.48550/arXiv.2405.13898}
  {10.48550/arXiv.2405.13898} (\bibinfo {year} {2024})\BibitemShut {NoStop}%
\bibitem [{\citenamefont {Alevras}\ \emph {et~al.}(2024)\citenamefont
  {Alevras}, \citenamefont {Metkar}, \citenamefont {Yamamoto}, \citenamefont
  {Kumar}, \citenamefont {Friedhoff}, \citenamefont {Park}, \citenamefont
  {Takeori}, \citenamefont {LaDue}, \citenamefont {David},\ and\ \citenamefont
  {Galda}}]{alevras:2024}%
  \BibitemOpen
  \bibfield  {author} {\bibinfo {author} {\bibfnamefont {D.}~\bibnamefont
  {Alevras}}, \bibinfo {author} {\bibfnamefont {M.}~\bibnamefont {Metkar}},
  \bibinfo {author} {\bibfnamefont {T.}~\bibnamefont {Yamamoto}}, \bibinfo
  {author} {\bibfnamefont {V.}~\bibnamefont {Kumar}}, \bibinfo {author}
  {\bibfnamefont {T.}~\bibnamefont {Friedhoff}}, \bibinfo {author}
  {\bibfnamefont {J.-E.}\ \bibnamefont {Park}}, \bibinfo {author}
  {\bibfnamefont {M.}~\bibnamefont {Takeori}}, \bibinfo {author} {\bibfnamefont
  {M.}~\bibnamefont {LaDue}}, \bibinfo {author} {\bibfnamefont
  {W.}~\bibnamefont {David}},\ and\ \bibinfo {author} {\bibfnamefont
  {A.}~\bibnamefont {Galda}},\ }\bibfield  {title} {\bibinfo {title} {m{RNA}
  secondary structure prediction using utility-scale quantum computers},\
  }\bibfield  {journal} {\bibinfo  {journal} {arXiv:2405.20328}\ }\href
  {https://doi.org/10.48550/arXiv.2405.20328} {10.48550/arXiv.2405.20328}
  (\bibinfo {year} {2024})\BibitemShut {NoStop}%
\bibitem [{\citenamefont {Sachdeva}\ \emph {et~al.}(2024)\citenamefont
  {Sachdeva}, \citenamefont {Hartnett}, \citenamefont {Maity}, \citenamefont
  {Marsh}, \citenamefont {Wang}, \citenamefont {Winick}, \citenamefont
  {Dougherty}, \citenamefont {Canuto}, \citenamefont {Chong}, \citenamefont
  {Hush}, \citenamefont {Mundada}, \citenamefont {Bentley}, \citenamefont
  {Biercuk},\ and\ \citenamefont {Baum}}]{sachdeva:2024}%
  \BibitemOpen
  \bibfield  {author} {\bibinfo {author} {\bibfnamefont {N.}~\bibnamefont
  {Sachdeva}}, \bibinfo {author} {\bibfnamefont {G.~S.}\ \bibnamefont
  {Hartnett}}, \bibinfo {author} {\bibfnamefont {S.}~\bibnamefont {Maity}},
  \bibinfo {author} {\bibfnamefont {S.}~\bibnamefont {Marsh}}, \bibinfo
  {author} {\bibfnamefont {Y.}~\bibnamefont {Wang}}, \bibinfo {author}
  {\bibfnamefont {A.}~\bibnamefont {Winick}}, \bibinfo {author} {\bibfnamefont
  {R.}~\bibnamefont {Dougherty}}, \bibinfo {author} {\bibfnamefont
  {D.}~\bibnamefont {Canuto}}, \bibinfo {author} {\bibfnamefont {Y.~Q.}\
  \bibnamefont {Chong}}, \bibinfo {author} {\bibfnamefont {M.}~\bibnamefont
  {Hush}}, \bibinfo {author} {\bibfnamefont {P.~S.}\ \bibnamefont {Mundada}},
  \bibinfo {author} {\bibfnamefont {C.~D.~B.}\ \bibnamefont {Bentley}},
  \bibinfo {author} {\bibfnamefont {M.~J.}\ \bibnamefont {Biercuk}},\ and\
  \bibinfo {author} {\bibfnamefont {Y.}~\bibnamefont {Baum}},\ }\bibfield
  {title} {\bibinfo {title} {{Q}uantum optimization using a 127-qubit
  gate-model {IBM} quantum computer can outperform quantum annealers for
  nontrivial binary optimization problems},\ }\bibfield  {journal} {\bibinfo
  {journal} {arXiv:2406.01743}\ }\href
  {https://doi.org/10.48550/arXiv.2406.01743} {10.48550/arXiv.2406.01743}
  (\bibinfo {year} {2024})\BibitemShut {NoStop}%
\bibitem [{cud(2024)}]{cudaq}%
  \BibitemOpen
  \href {https://github.com/NVIDIA/cuda-quantum} {\bibinfo {title}
  {https://github.com/nvidia/cuda-quantum}} (\bibinfo {year}
  {2024})\BibitemShut {NoStop}%
\bibitem [{\citenamefont {Krekel}\ \emph {et~al.}()\citenamefont {Krekel},
  \citenamefont {Oliveira}, \citenamefont {Pfannschmidt}, \citenamefont
  {Bruynooghe}, \citenamefont {Laugher},\ and\ \citenamefont
  {Bruhin}}]{pytest}%
  \BibitemOpen
  \bibfield  {author} {\bibinfo {author} {\bibfnamefont {H.}~\bibnamefont
  {Krekel}}, \bibinfo {author} {\bibfnamefont {B.}~\bibnamefont {Oliveira}},
  \bibinfo {author} {\bibfnamefont {R.}~\bibnamefont {Pfannschmidt}}, \bibinfo
  {author} {\bibfnamefont {F.}~\bibnamefont {Bruynooghe}}, \bibinfo {author}
  {\bibfnamefont {B.}~\bibnamefont {Laugher}},\ and\ \bibinfo {author}
  {\bibfnamefont {F.}~\bibnamefont {Bruhin}},\ }\href
  {https://github.com/pytest-dev/pytest} {\bibinfo {title} {pytest
  x.y}}\BibitemShut {NoStop}%
\bibitem [{Note1()}]{Note1}%
  \BibitemOpen
  \bibinfo {note} {Because we execute tests in subprocesses to implement a
  timeout mechanism, some failures can kill the subprocess but otherwise not
  affect the remaining tests. These are considered \protect \texttt {FAILED}
  per the definition here.}\BibitemShut {Stop}%
\bibitem [{\citenamefont {Cross}\ \emph {et~al.}(2019)\citenamefont {Cross},
  \citenamefont {Bishop}, \citenamefont {Sheldon}, \citenamefont {Nation},\
  and\ \citenamefont {Gambetta}}]{cross:2019}%
  \BibitemOpen
  \bibfield  {author} {\bibinfo {author} {\bibfnamefont {A.~W.}\ \bibnamefont
  {Cross}}, \bibinfo {author} {\bibfnamefont {L.~S.}\ \bibnamefont {Bishop}},
  \bibinfo {author} {\bibfnamefont {S.}~\bibnamefont {Sheldon}}, \bibinfo
  {author} {\bibfnamefont {P.~D.}\ \bibnamefont {Nation}},\ and\ \bibinfo
  {author} {\bibfnamefont {J.~M.}\ \bibnamefont {Gambetta}},\ }\bibfield
  {title} {\bibinfo {title} {{V}alidating quantum computers using randomized
  model circuits},\ }\href {https://doi.org/10.1103/PhysRevA.100.032328}
  {\bibfield  {journal} {\bibinfo  {journal} {Phys. Rev. A}\ }\textbf {\bibinfo
  {volume} {100}},\ \bibinfo {pages} {032328} (\bibinfo {year}
  {2019})}\BibitemShut {NoStop}%
\bibitem [{\citenamefont {Bravyi}\ and\ \citenamefont {Maslov}(2021)}]{BM2021}%
  \BibitemOpen
  \bibfield  {author} {\bibinfo {author} {\bibfnamefont {S.}~\bibnamefont
  {Bravyi}}\ and\ \bibinfo {author} {\bibfnamefont {D.}~\bibnamefont
  {Maslov}},\ }\bibfield  {title} {\bibinfo {title} {Hadamard-free circuits
  expose the structure of the clifford group},\ }\href
  {https://doi.org/10.1109/TIT.2021.3081415} {\bibfield  {journal} {\bibinfo
  {journal} {IEEE Transactions on Information Theory}\ }\textbf {\bibinfo
  {volume} {67}},\ \bibinfo {pages} {4546} (\bibinfo {year}
  {2021})}\BibitemShut {NoStop}%
\bibitem [{\citenamefont {Bravyi}\ \emph {et~al.}(2021)\citenamefont {Bravyi},
  \citenamefont {Shaydulin}, \citenamefont {Hu},\ and\ \citenamefont
  {Maslov}}]{Bravyi2021cliffordcircuit}%
  \BibitemOpen
  \bibfield  {author} {\bibinfo {author} {\bibfnamefont {S.}~\bibnamefont
  {Bravyi}}, \bibinfo {author} {\bibfnamefont {R.}~\bibnamefont {Shaydulin}},
  \bibinfo {author} {\bibfnamefont {S.}~\bibnamefont {Hu}},\ and\ \bibinfo
  {author} {\bibfnamefont {D.}~\bibnamefont {Maslov}},\ }\bibfield  {title}
  {\bibinfo {title} {Clifford {C}ircuit {O}ptimization with {T}emplates and
  {S}ymbolic {P}auli {G}ates},\ }\href
  {https://doi.org/10.22331/q-2021-11-16-580} {\bibfield  {journal} {\bibinfo
  {journal} {{Quantum}}\ }\textbf {\bibinfo {volume} {5}},\ \bibinfo {pages}
  {580} (\bibinfo {year} {2021})}\BibitemShut {NoStop}%
\bibitem [{\citenamefont {Knill}(2004)}]{knill:2004}%
  \BibitemOpen
  \bibfield  {author} {\bibinfo {author} {\bibfnamefont {E.}~\bibnamefont
  {Knill}},\ }\bibfield  {title} {\bibinfo {title} {{F}ault-{T}olerant
  {P}ostselected {Q}uantum {C}omputation: {T}hreshold {A}nalysis},\ }\bibfield
  {journal} {\bibinfo  {journal} {arXiv:0404104}\ }\href
  {https://doi.org/10.48550/arXiv.quant-ph/0404104}
  {10.48550/arXiv.quant-ph/0404104} (\bibinfo {year} {2004})\BibitemShut
  {NoStop}%
\bibitem [{\citenamefont {Wallman}\ and\ \citenamefont
  {Emerson}(2016)}]{wallman:2016}%
  \BibitemOpen
  \bibfield  {author} {\bibinfo {author} {\bibfnamefont {J.~J.}\ \bibnamefont
  {Wallman}}\ and\ \bibinfo {author} {\bibfnamefont {J.}~\bibnamefont
  {Emerson}},\ }\bibfield  {title} {\bibinfo {title} {{N}oise tailoring for
  scalable quantum computation via randomized compiling},\ }\href
  {https://doi.org/10.1103/PhysRevA.94.052325} {\bibfield  {journal} {\bibinfo
  {journal} {Phys. Rev. A}\ }\textbf {\bibinfo {volume} {94}},\ \bibinfo
  {pages} {052325} (\bibinfo {year} {2016})}\BibitemShut {NoStop}%
\bibitem [{bra()}]{braketprovider}%
  \BibitemOpen
  \href {https://github.com/qiskit-community/qiskit-braket-provider} {\bibinfo
  {title}
  {https://github.com/qiskit-community/qiskit-braket-provider}}\BibitemShut
  {NoStop}%
\bibitem [{\citenamefont {Fleming}\ and\ \citenamefont
  {Wallace}(1986)}]{stats}%
  \BibitemOpen
  \bibfield  {author} {\bibinfo {author} {\bibfnamefont {P.~J.}\ \bibnamefont
  {Fleming}}\ and\ \bibinfo {author} {\bibfnamefont {J.~J.}\ \bibnamefont
  {Wallace}},\ }\bibfield  {title} {\bibinfo {title} {How not to lie with
  statistics: the correct way to summarize benchmark results},\ }\href
  {https://doi.org/10.1145/5666.5673} {\bibfield  {journal} {\bibinfo
  {journal} {Commun. ACM}\ }\textbf {\bibinfo {volume} {29}},\ \bibinfo {pages}
  {218} (\bibinfo {year} {1986})}\BibitemShut {NoStop}%
\bibitem [{\citenamefont {Treinish}\ \emph {et~al.}(2022)\citenamefont
  {Treinish}, \citenamefont {Carvalho}, \citenamefont {Tsilimigkounakis},\ and\
  \citenamefont {S{\'a}}}]{rustworkx}%
  \BibitemOpen
  \bibfield  {author} {\bibinfo {author} {\bibfnamefont {M.}~\bibnamefont
  {Treinish}}, \bibinfo {author} {\bibfnamefont {I.}~\bibnamefont {Carvalho}},
  \bibinfo {author} {\bibfnamefont {G.}~\bibnamefont {Tsilimigkounakis}},\ and\
  \bibinfo {author} {\bibfnamefont {N.}~\bibnamefont {S{\'a}}},\ }\bibfield
  {title} {\bibinfo {title} {rustworkx: A high-performance graph library for
  python},\ }\href {https://doi.org/10.21105/joss.03968} {\bibfield  {journal}
  {\bibinfo  {journal} {Journal of Open Source Software}\ }\textbf {\bibinfo
  {volume} {7}},\ \bibinfo {pages} {3968} (\bibinfo {year} {2022})}\BibitemShut
  {NoStop}%
\bibitem [{Note2()}]{Note2}%
  \BibitemOpen
  \bibinfo {note} {This data was added after the initial version of Benchpress,
  and is not included in the published results.}\BibitemShut {Stop}%
\bibitem [{\citenamefont {L{\"o}tstedt}\ and\ \citenamefont
  {Yamanouchi}(2024)}]{lotstedt:2024}%
  \BibitemOpen
  \bibfield  {author} {\bibinfo {author} {\bibfnamefont {E.}~\bibnamefont
  {L{\"o}tstedt}}\ and\ \bibinfo {author} {\bibfnamefont {K.}~\bibnamefont
  {Yamanouchi}},\ }\bibfield  {title} {\bibinfo {title} {Comparison of current
  quantum devices for quantum computing of heisenberg spin chain dynamics},\
  }\href {https://doi.org/https://doi.org/10.1016/j.cplett.2023.140975}
  {\bibfield  {journal} {\bibinfo  {journal} {Chemical Physics Letters}\
  }\textbf {\bibinfo {volume} {836}},\ \bibinfo {pages} {140975} (\bibinfo
  {year} {2024})}\BibitemShut {NoStop}%
\bibitem [{mem()}]{memray}%
  \BibitemOpen
  \href {https://github.com/bloomberg/memray} {\bibinfo {title}
  {https://github.com/bloomberg/memray}}\BibitemShut {NoStop}%
\bibitem [{ben({\natexlab{a}})}]{benchpress}%
  \BibitemOpen
  \href {https://github.com/Qiskit/benchpress} {\bibinfo {title}
  {https://github.com/qiskit/benchpress}} ({\natexlab{a}})\BibitemShut
  {NoStop}%
\bibitem [{ben({\natexlab{b}})}]{benchmark}%
  \BibitemOpen
  \href {https://github.com/ionelmc/pytest-benchmark/} {\bibinfo {title}
  {https://github.com/ionelmc/pytest-benchmark/}} ({\natexlab{b}})\BibitemShut
  {NoStop}%
\end{thebibliography}%


\begin{thebibliography}{3}%
\makeatletter
\providecommand \@ifxundefined [1]{%
 \@ifx{#1\undefined}
}%
\providecommand \@ifnum [1]{%
 \ifnum #1\expandafter \@firstoftwo
 \else \expandafter \@secondoftwo
 \fi
}%
\providecommand \@ifx [1]{%
 \ifx #1\expandafter \@firstoftwo
 \else \expandafter \@secondoftwo
 \fi
}%
\providecommand \natexlab [1]{#1}%
\providecommand \enquote  [1]{``#1''}%
\providecommand \bibnamefont  [1]{#1}%
\providecommand \bibfnamefont [1]{#1}%
\providecommand \citenamefont [1]{#1}%
\providecommand \href@noop [0]{\@secondoftwo}%
\providecommand \href [0]{\begingroup \@sanitize@url \@href}%
\providecommand \@href[1]{\@@startlink{#1}\@@href}%
\providecommand \@@href[1]{\endgroup#1\@@endlink}%
\providecommand \@sanitize@url [0]{\catcode `\\12\catcode `\$12\catcode
  `\&12\catcode `\#12\catcode `\^12\catcode `\_12\catcode `\%12\relax}%
\providecommand \@@startlink[1]{}%
\providecommand \@@endlink[0]{}%
\providecommand \url  [0]{\begingroup\@sanitize@url \@url }%
\providecommand \@url [1]{\endgroup\@href {#1}{\urlprefix }}%
\providecommand \urlprefix  [0]{URL }%
\providecommand \Eprint [0]{\href }%
\providecommand \doibase [0]{https://doi.org/}%
\providecommand \selectlanguage [0]{\@gobble}%
\providecommand \bibinfo  [0]{\@secondoftwo}%
\providecommand \bibfield  [0]{\@secondoftwo}%
\providecommand \translation [1]{[#1]}%
\providecommand \BibitemOpen [0]{}%
\providecommand \bibitemStop [0]{}%
\providecommand \bibitemNoStop [0]{.\EOS\space}%
\providecommand \EOS [0]{\spacefactor3000\relax}%
\providecommand \BibitemShut  [1]{\csname bibitem#1\endcsname}%
\let\auto@bib@innerbib\@empty
\bibitem [{\citenamefont {Amy}(2019)}]{amy:2019}%
  \BibitemOpen
  \bibfield  {author} {\bibinfo {author} {\bibfnamefont {M.}~\bibnamefont
  {Amy}},\ }\bibfield  {title} {\bibinfo {title} {Towards large-scale
  functional verification of universal quantum circuits},\ }in\ \href
  {https://doi.org/10.4204/EPTCS.287.1} {\emph {\bibinfo {booktitle}
  {Proceedings of the 15th International Conference on Quantum Physics and
  Logic}}}\ (\bibinfo {year} {2019})\ pp.\ \bibinfo {pages} {1--21}\BibitemShut
  {NoStop}%
\bibitem [{\citenamefont {Sawaya}\ \emph {et~al.}(2023)\citenamefont {Sawaya},
  \citenamefont {Marti-Dafcik}, \citenamefont {Ho}, \citenamefont {Tabor},
  \citenamefont {Bernal~Neira}, \citenamefont {Magann}, \citenamefont
  {Premaratne}, \citenamefont {Dubey}, \citenamefont {Matsuura}, \citenamefont
  {Bishop}, \citenamefont {De~Jong}, \citenamefont {Benjamin}, \citenamefont
  {Parekh}, \citenamefont {Tubman}, \citenamefont {Klymko},\ and\ \citenamefont
  {Camps}}]{sawaya:2023}%
  \BibitemOpen
  \bibfield  {author} {\bibinfo {author} {\bibfnamefont {N.~P.}\ \bibnamefont
  {Sawaya}}, \bibinfo {author} {\bibfnamefont {D.}~\bibnamefont
  {Marti-Dafcik}}, \bibinfo {author} {\bibfnamefont {Y.}~\bibnamefont {Ho}},
  \bibinfo {author} {\bibfnamefont {D.~P.}\ \bibnamefont {Tabor}}, \bibinfo
  {author} {\bibfnamefont {D.~E.}\ \bibnamefont {Bernal~Neira}}, \bibinfo
  {author} {\bibfnamefont {A.~B.}\ \bibnamefont {Magann}}, \bibinfo {author}
  {\bibfnamefont {S.}~\bibnamefont {Premaratne}}, \bibinfo {author}
  {\bibfnamefont {P.}~\bibnamefont {Dubey}}, \bibinfo {author} {\bibfnamefont
  {A.}~\bibnamefont {Matsuura}}, \bibinfo {author} {\bibfnamefont
  {N.}~\bibnamefont {Bishop}}, \bibinfo {author} {\bibfnamefont {W.~A.}\
  \bibnamefont {De~Jong}}, \bibinfo {author} {\bibfnamefont {S.}~\bibnamefont
  {Benjamin}}, \bibinfo {author} {\bibfnamefont {O.~D.}\ \bibnamefont
  {Parekh}}, \bibinfo {author} {\bibfnamefont {N.~M.}\ \bibnamefont {Tubman}},
  \bibinfo {author} {\bibfnamefont {K.}~\bibnamefont {Klymko}},\ and\ \bibinfo
  {author} {\bibfnamefont {D.}~\bibnamefont {Camps}},\ }\bibfield  {title}
  {\bibinfo {title} {{H}am{L}ib: {A} {L}ibrary of {H}amiltonians for
  {B}enchmarking {Q}uantum {A}lgorithms and {H}ardware},\ }in\ \href
  {https://doi.org/10.1109/QCE57702.2023.10296} {\emph {\bibinfo {booktitle}
  {2023 IEEE International Conference on Quantum Computing and Engineering
  (QCE)}}},\ Vol.~\bibinfo {volume} {2}\ (\bibinfo {year} {2023})\ p.\ \bibinfo
  {pages} {389}\BibitemShut {NoStop}%
\bibitem [{\citenamefont {Li}\ \emph {et~al.}(2023)\citenamefont {Li},
  \citenamefont {Stein}, \citenamefont {Krishnamoorthy},\ and\ \citenamefont
  {Ang}}]{li:2023}%
  \BibitemOpen
  \bibfield  {author} {\bibinfo {author} {\bibfnamefont {A.}~\bibnamefont
  {Li}}, \bibinfo {author} {\bibfnamefont {S.}~\bibnamefont {Stein}}, \bibinfo
  {author} {\bibfnamefont {S.}~\bibnamefont {Krishnamoorthy}},\ and\ \bibinfo
  {author} {\bibfnamefont {J.}~\bibnamefont {Ang}},\ }\bibfield  {title}
  {\bibinfo {title} {Qasmbench: A low-level quantum benchmark suite for nisq
  evaluation and simulation},\ }\bibfield  {journal} {\bibinfo  {journal} {ACM
  Transactions on Quantum Computing}\ }\textbf {\bibinfo {volume} {4}},\ \href
  {https://doi.org/10.1145/3550488} {10.1145/3550488} (\bibinfo {year}
  {2023})\BibitemShut {NoStop}%
\end{thebibliography}%
\end{document}


\title{Supplemental Information: Benchmarking the performance of quantum computing software}

\author{Paul D. Nation}
\email[E-mail: ]{paul.nation@ibm.com}
\author{Abdullah Ash Saki}
\affiliation{IBM Quantum, IBM T. J. Watson Research Center, Yorktown Heights, NY, 10598 USA}
\author{Sebastian Brandhofer}
\affiliation{IBM Quantum, IBM Germany Research \& Development, B\"{o}blingen Germany}
\author{Luciano Bello}
\affiliation{IBM Quantum, IBM Research Europe, Zurich Switzerland}
\author{Shelly Garion}
\affiliation{IBM Quantum, IBM Research Israel, Haifa 3498825, Israel}
\author{Matthew Treinish}
\author{Ali Javadi-Abhari}
\affiliation{IBM Quantum, IBM T. J. Watson Research Center, Yorktown Heights, NY, 10598 USA}

\date{\today}

\maketitle

\section{Circuit manipulation decomposition test gate counts and depths}\label{sec:decomp}

\begin{figure*}[h]
\centering
\includegraphics[width=13cm]{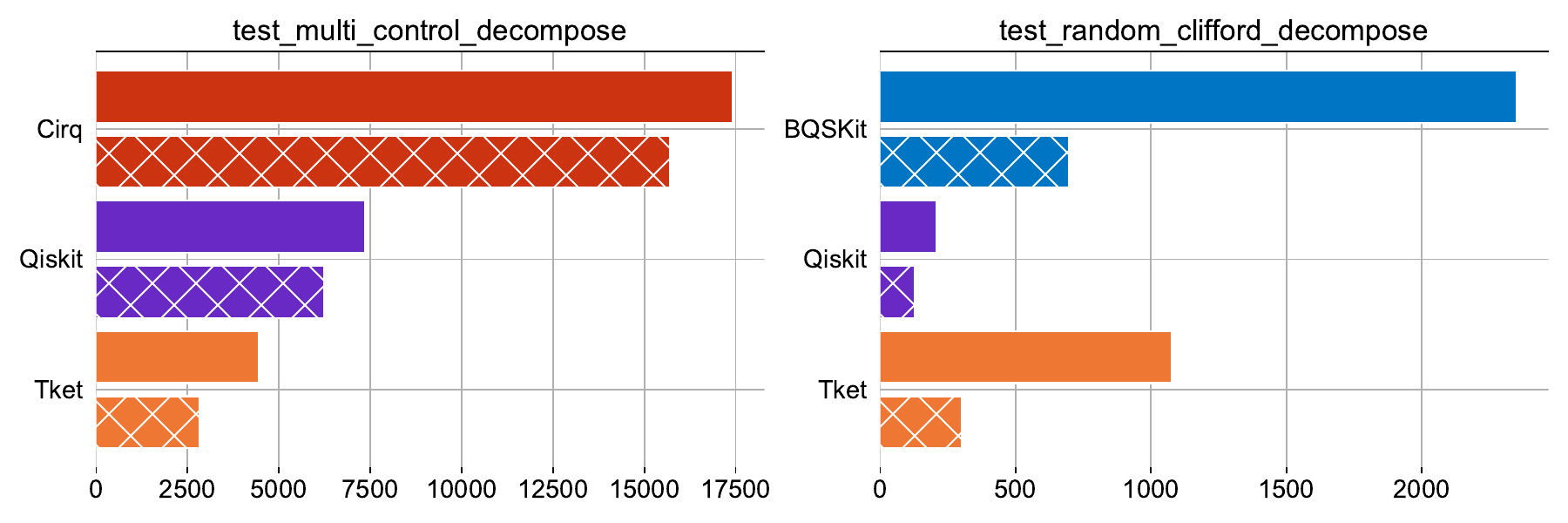}
\caption{Two-qubit gate count (solid) and depth (hatched) for successful decomposition tests evaluating SDK circuit manipulation performance. Lower values are better.  Only SDKs with successful test outputs are listed.}
\label{fig:stats}
\end{figure*}

Figure~(2) of the main text gives timing comparisons for quantum circuit construction and manipulation tests.  Within the latter set there are two decomposition tests that measure not only the time it takes to complete the routines, but also gauge the synthesis capabilities of each SDK.  As such, the two-qubit gate count and depth are important quantities of interest.  In Fig.~(\ref{fig:stats}) we present these values for each of the two decomposition tests, \texttt{multi\_control\_decompose} and \texttt{random\_clifford\_decomposition}, in Benchpress.

\clearpage
\section{Circuit transpilation results per test grouping}\label{sec:groupings}

In the main text, Fig.~(3) and Tbl.~(2) give results for the entirety of the Benchpress transpilation tests.  In this section we give a more granular view, and detail the results in isolation for each of the open-source data sets used in this work, i.e. the Feynman \cite{amy:2019}, Hamlib \cite{sawaya:2023}, and QasmBench \cite{li:2023} libraries.

\subsection{Results for circuits from Feynman suite}\label{sec:feynman}

\begin{figure*}[h]
\centering
\includegraphics[width=14cm]{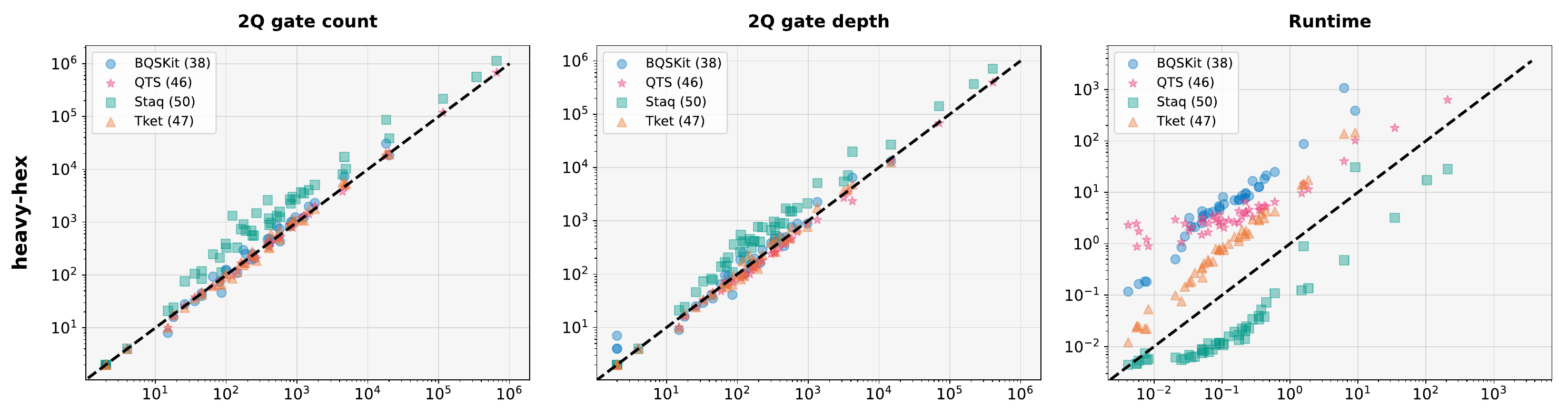}
\caption{Subset of results generated from the device transpilation tests in Benchpress for circuits generated from the Feynman test suite \cite{amy:2019}.  Rows indicate the topology used in the tests, whereas columns label the reported metric.  For brevity, only the top row of plots is labeled.  The target topology for the device transpilation tests is ``heavy-hex".  The total number of passed tests in each data set is given in the legends.}
\label{fig:feynman_results}
\end{figure*}

\begin{table*}[h]
\centering
\scriptsize
\ra{1.3}
\begin{tabular}{@{}rccccccccccccccc@{}}\toprule
& \multicolumn{4}{c}{\textbf{BQSKit}} &  \multicolumn{4}{c}{\textbf{QTS}}  & \multicolumn{4}{c}{\textbf{Staq}}  & \multicolumn{3}{c}{\textbf{Tket}}\\
\cmidrule{2-4} \cmidrule{6-8} \cmidrule{10-12} \cmidrule{14-16}

& ~2Q gates~ & ~2Q depth~ & ~time~ && ~2Q gates~ & ~2Q depth~ & ~time~ && ~2Q gates~ & ~2Q depth~ & ~time~ && ~2Q gates~ & ~2Q depth~ & ~time~\\ \midrule

\textbf{heavy-hex} & $1.07$/$1.07$ & $1.12$/$1.10$ & $47.3$/$46.6$ && 
$0.96$/$1.0$ & $0.83$/$0.83$ & $28.2$/$24.3$ && 
$2.56$/$2.73$ & $2.16$/$2.21$ & $0.19$/$0.14$ && 
$0.98$/$1.9$ & $0.97$/$1.0$ & $6.89$/$7.27$ \\

\bottomrule
\end{tabular}
\caption{Geometric mean / median values for SDK performance metrics for circuits from the Feynman test suite.  Results are normalized to their corresponding Qiskit values. Feynman tests are device transpilation tests and have been run over only a heavy-hex topology.}
\label{tbl:fyenman}
\end{table*}

\clearpage
\subsection{Results for circuits from HamLib library}\label{sec:hamlib}

\begin{figure*}[th]
\centering
\includegraphics[width=13cm]{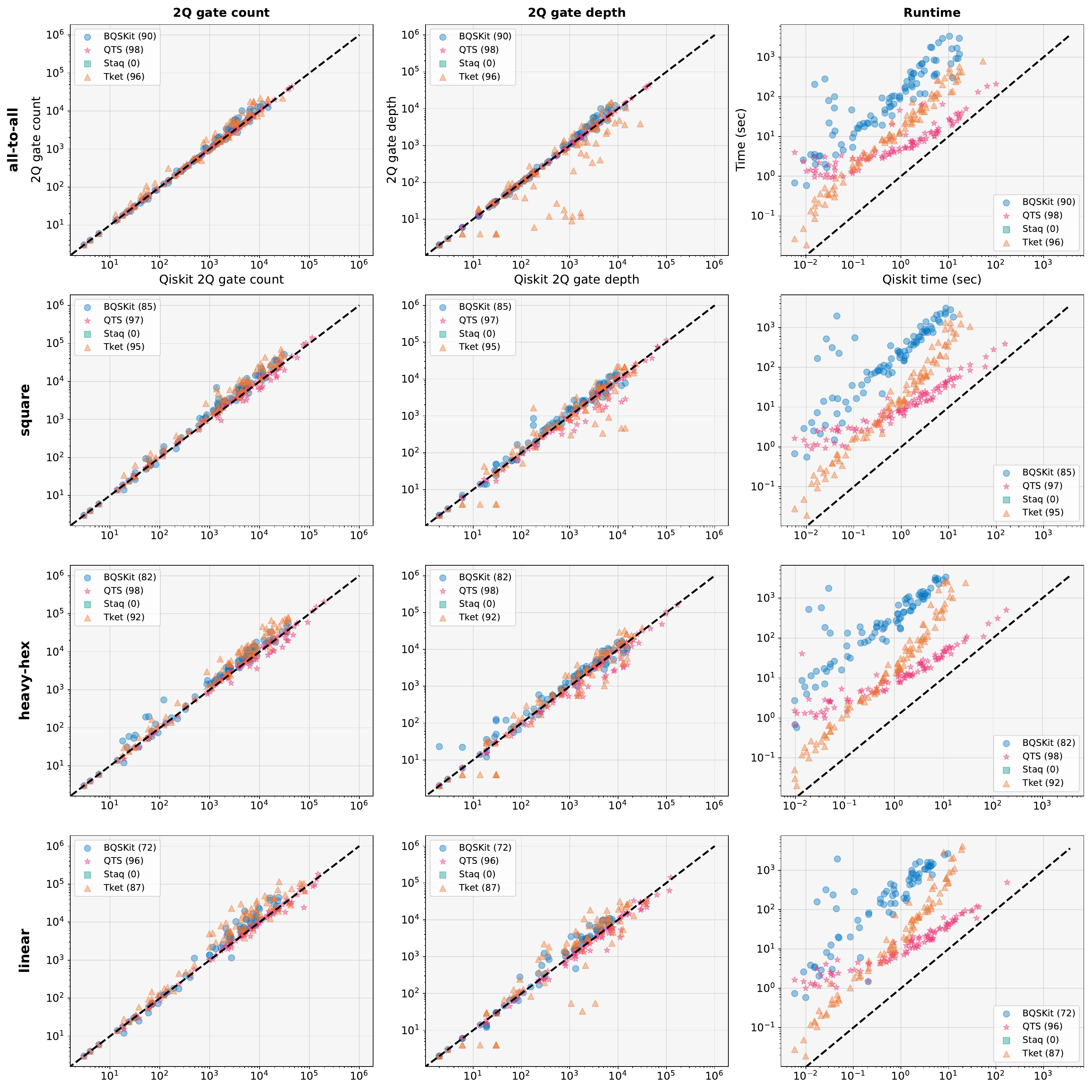}
\caption{Subset of results generated from the device- and abstract transpilation tests in Benchpress for circuits generated from Hamiltonian operators in the Hamlib library \cite{sawaya:2023}.  Rows indicate the topology used in the tests, whereas columns label the reported metric.  For brevity, only the top row of plots is labeled.  The total number of passed tests in each data set is given in the legends.}
\label{fig:hamlib_results}
\end{figure*}

\begin{table*}[h]
\centering
\scriptsize
\ra{1.3}
\begin{tabular}{@{}rccccccccccc@{}}\toprule
& \multicolumn{4}{c}{\textbf{BQSKit}} &  \multicolumn{4}{c}{\textbf{QTS}}  & \multicolumn{3}{c}{\textbf{Tket}}\\
\cmidrule{2-4} \cmidrule{6-8} \cmidrule{10-12}

& ~2Q gates~ & ~2Q depth~ & ~time~ && ~2Q gates~ & ~2Q depth~ & ~time~ && ~2Q gates~ & ~2Q depth~ & ~time~\\ \midrule

\textbf{All tests} & $1.28$/$1.26$ & $1.23$/$1.21$ & $288$/$276$ && 
$0.97$/$1.0$ & $0.88$/$1.0$ & $10.6$/$7.71$ && 
$1.51$/$1.42$ & $0.93$/$1.16$ & $19.0$/$17.0$ \\

\textbf{all-to-all} & $1.10$/$1.0$ & $1.10$/$1.0$ & $157$/$133$ && 
$1.0$/$1.0$ & $1.0$/$1.0$ & $9.26$/$6.52$ && 
$1.22$/$1.09$ & $0.57$/$1.00$ & $16.4$/$16.8$ \\

\textbf{square} & $1.32$/$1.31$ & $1.21$/$1.21$ & $287$/$247$ && 
$0.99$/$1.03$ & $0.84$/$0.96$ & $10.9$/$8.03$ && 
$1.49$/$1.38$ & $0.93$/$1.10$ & $18.9$/$16.7$ \\

\textbf{heavy-hex} & $1.36$/$1.33$ & $1.29$/$1.25$ & $377$/$321$ &&
$0.96$/$1.0$ & $0.85$/$0.94$ & $11.3$/$7.87$ && 
$1.58$/$1.55$ & $1.09$/$1.25$ & $18.0$/$15.9$ \\

\textbf{linear} & $1.34$/$1.33$ & $1.28$/$1.24$ & $352$/$323$ && 
$0.96$/$1.0$ & $0.87$/$0.96$ & $10.6$/$7.21$ && 
$1.77$/$1.56$ & $1.16$/$1.47$ & $24.8$/$22.9$ \\

\bottomrule
\end{tabular}
\caption{Geometric mean / median values for SDK performance metrics for circuits derived from the Hamlib library of Hamiltonians.  Results are normalized to their corresponding Qiskit values.  As the circuits cannot be expressed in QASM form, they are not compatible inputs for the Staq compiler.}
\label{tbl:hamlib}
\end{table*}

\clearpage
\subsection{Results for circuits from QasmBench suite}\label{sec:qasmbench}

\begin{figure*}[th]
\centering
\includegraphics[width=13cm]{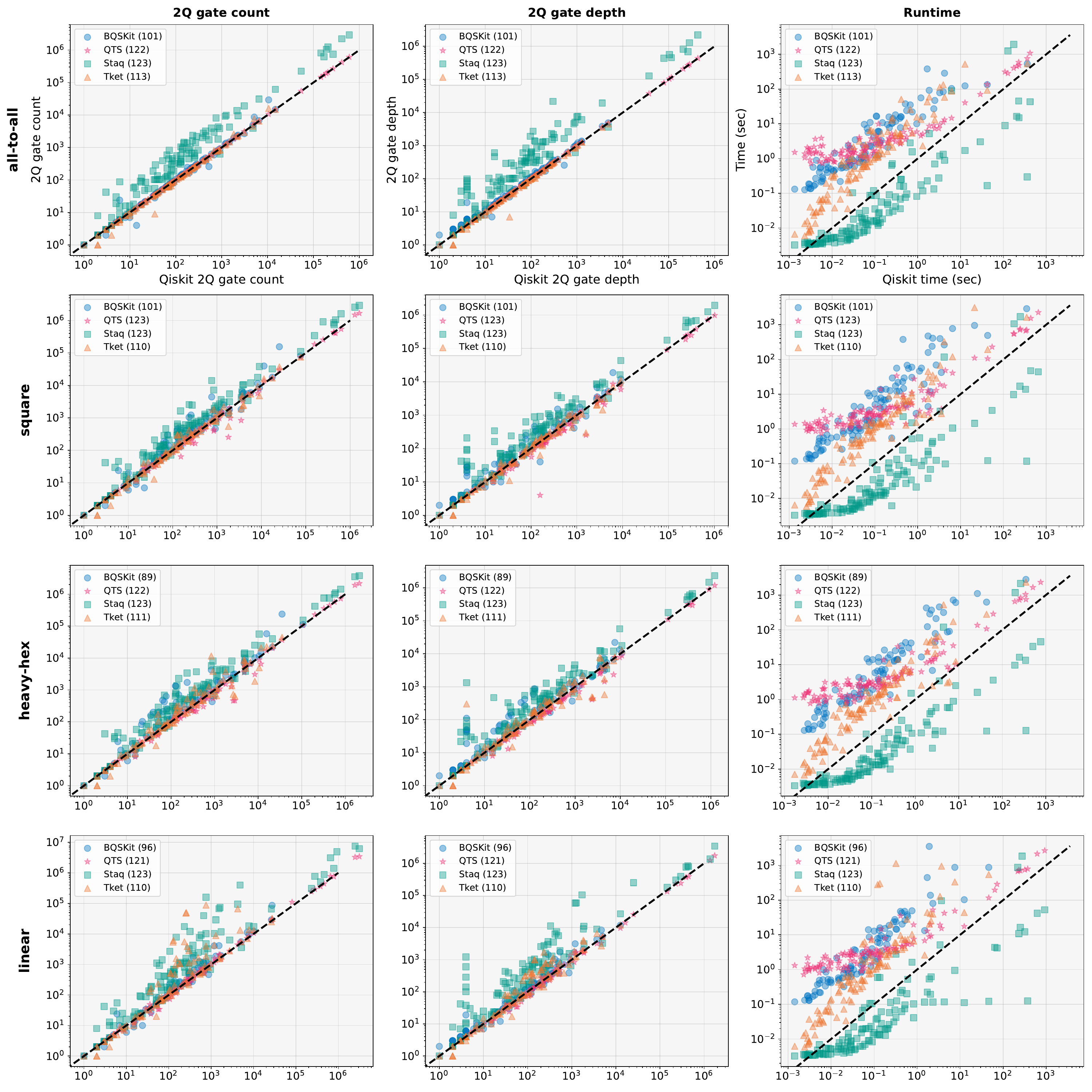}
\caption{Subset of results generated from the abstract transpilation tests in Benchpress for circuits generated from the QasmBench test suite \cite{li:2023}.  Rows indicate the topology used in the tests, whereas columns label the reported metric.  For brevity, only the top row of plots is labeled.  The total number of passed tests in each data set is given in the legends.}
\label{fig:qasmbench_results}
\end{figure*}

\begin{table*}[h]
\centering
\scriptsize
\ra{1.3}
\begin{tabular}{@{}rccccccccccccccc@{}}\toprule
& \multicolumn{4}{c}{\textbf{BQSKit}} &  \multicolumn{4}{c}{\textbf{QTS}}  & \multicolumn{4}{c}{\textbf{Staq}}  & \multicolumn{3}{c}{\textbf{Tket}}\\
\cmidrule{2-4} \cmidrule{6-8} \cmidrule{10-12} \cmidrule{14-16}

& ~2Q gates~ & ~2Q depth~ & ~time~ && ~2Q gates~ & ~2Q depth~ & ~time~ && ~2Q gates~ & ~2Q depth~ & ~time~ && ~2Q gates~ & ~2Q depth~ & ~time~\\ \midrule

\textbf{All tests} & $1.25$/$1.04$ & $1.33$/$1.17$ & $43.9$/$43.7$ && 
$0.95$/$1.0$ & $0.91$/$1.0$ & $31.8$/$30.0$ && 
$2.83$/$2.44$ & $2.99$/$2.43$ & $0.27$/$0.25$ && 
$1.17$/$1.0$ & $1.05$/$1.0$ & $10.2$/$10.2$ \\

\textbf{all-to-all} & $1.0$/$1.0$ & $1.13$/$1.09$ & $38.9$/$41.0$ && 
$1.0$/$1.0$ & $1.0$/$1.0$ & $31.8$/$29.0$ && 
$3.35$/$3.76$ & $4.02$/$4.07$ & $0.36$/$0.34$ && 
$0.97$/$1.00$ & $0.95$/$1.00$ & $12.5$/$11.3$ \\

\textbf{square} & $1.31$/$1.15$ & $1.39$/$1.25$ & $44.1$/$45.6$ && 
$0.94$/$1.0$ & $0.86$/$0.98$ & $29.6$/$29.5$ && 
$2.15$/$2.13$ & $2.36$/$2.14$ & $0.23$/$0.19$ && 
$1.01$/$1.0$ & $0.95$/$1.00$ & $8.65$/$9.39$ \\

\textbf{heavy-hex} & $1.64$/$1.30$ & $1.70$/$1.33$ & $43.8$/$41.4$ &&
 $0.89$/$1.0$ & $0.83$/$0.90$ & $29.1$/$27.7$ && 
 $2.27$/$2.09$ & $2.31$/$1.77$ & $0.21$/$0.19$ && 
 $1.10$/$1.0$ & $1.04$/$1.0$ & $7.74$/$8.80$ \\

\textbf{linear} & $1.15$/$1.0$ & $1.22$/$1.15$ & $49.7$/$45.9$ && 
$1.0$/$1.0$ & $0.94$/$1.0$ & $37.5$/$37.1$ && 
$3.92$/$3.48$ & $3.66$/$2.75$ & $0.30$/$0.31$ && 
$1.76$/$1.0$ & $1.26$/$1.0$ & $12.9$/$10.7$ \\

\bottomrule
\end{tabular}
\caption{Geometric mean / median values for SDK performance metrics for circuits from the QasmBench test suite.  Results are normalized to their corresponding Qiskit values.}
\label{tbl:qasmbench}
\end{table*}

\newpage
\bibliography{refs}